\documentclass[manuscript]{acmart}

\AtBeginDocument{%
  \providecommand\BibTeX{{%
    \normalfont B\kern-0.5em{\scshape i\kern-0.25em b}\kern-0.8em\TeX}}}

\setcopyright{acmlicensed}
\copyrightyear{2025}
\acmYear{2025}
\usepackage{graphicx}
\usepackage{pdflscape}  % for landscape pages

\begin{document}

\title[LLM-Powered AI Tutors for DHH Learners]{LLM-Powered AI Tutors with Personas for d/Deaf and Hard-of-Hearing Online Learners}

\author{Haocong Cheng}
\email{haocong2@illinois.edu}
\affiliation{
  \institution{School of Information Sciences, University of Illinois Urbana-Champaign}
  \city{Champaign}
  \state{Illinois}
  \country{USA}
}

\author{Si Chen}
\email{schen34@nd.edu}
\orcid{0000-0002-0640-6883}
\affiliation{%
  \institution{University of Notre Dame}
  \city{Notre Dame}
  \state{Indiana}
  \country{USA}
}
\affiliation{%
  \institution{School of Information Sciences, University of Illinois Urbana-Champaign}
  \city{Champaign}
  \state{Illinois}
  \country{USA}
}

\author{Christopher Perdriau}
\email{chp5@illinois.edu}
\affiliation{
  \institution{Computer Science, University of Illinois Urbana-Champaign}
  \city{Champaign}
  \state{Illinois}
  \country{USA}
}

\author{Shriya Mokkapati}
\email{shriyam3@illinois.edu}
\affiliation{
  \institution{School of Information Sciences, University of Illinois Urbana-Champaign}
  \city{Champaign}
  \state{Illinois}
  \country{USA}
}

\author{Yun Huang}
\email{yunhuang@illinois.edu}
% \authornote{Corresponding author}
\affiliation{
  \institution{School of Information Sciences, University of Illinois Urbana-Champaign}
  \city{Champaign}
  \state{Illinois}
  \country{USA}
}

\begin{abstract}
Intelligent tutoring systems (ITS) using artificial intelligence (AI) technology have shown promise in supporting learners with diverse abilities. Large language models (LLMs) provide new opportunities to incorporate personas to AI-based tutors and support dynamic interactive dialogue. This paper explores how DHH learners interact with LLM-powered AI tutors with different experiences in DHH education as personas to identify their accessibility preferences. A user study with 16 DHH participants showed that they asked DHH-related questions based on background information and evaluated the AI tutors' cultural knowledge of the DHH communities in their responses. Participants suggested providing more transparency in each AI tutor's position within the DHH community. Participants also pointed out the lack of support in the multimodality of sign language in current LLMs. We discuss design implications to support the diverse needs in interaction between DHH users and the LLMs, such as offering supports in tuning language styles of LLMs.
  % 149 words. Word limit 150 words.
\end{abstract}

\begin{CCSXML}
<ccs2012>
   <concept>
       <concept_id>10003120.10011738.10011774</concept_id>
       <concept_desc>Human-centered computing~Accessibility design and evaluation methods</concept_desc>
       <concept_significance>500</concept_significance>
       </concept>
   <concept>
       <concept_id>10003456.10010927.10003616</concept_id>
       <concept_desc>Social and professional topics~People with disabilities</concept_desc>
       <concept_significance>500</concept_significance>
       </concept>
   <concept>
       <concept_id>10010405.10010489.10010495</concept_id>
       <concept_desc>Applied computing~E-learning</concept_desc>
       <concept_significance>500</concept_significance>
       </concept>
 </ccs2012>
\end{CCSXML}

\ccsdesc[500]{Human-centered computing~Accessibility design and evaluation methods}
\ccsdesc[500]{Social and professional topics~People with disabilities}
\ccsdesc[500]{Applied computing~E-learning}
\keywords{Large-Language Models, d/Deaf and hard of hearing, Personas, Online Learning, Intelligent tutoring systems}


\maketitle

\section{Introduction}

In the United States, about two per 1,000 children have some level of hearing loss in one or both ears \cite{gaffney2010identifying}, and one out of 8 people (13\%) aged 12 or above has hearing loss in both ears \cite{lin2011hearing}. As one of the diverse ability populations, there are educational technologies designed to support d/Deaf and Hard-of-Hearing (DHH) learners to address their accessibility needs. Various AI technologies have been explored to support DHH learners through automated caption generation \cite{butler2019exploration, kushalnagar2012readability, lasecki2014helping}, emotional experience \cite{chen2024comparison}, and sign language translation \cite{papastratis2021artificial, bragg2019sign}. 

\begin{figure*}[t]
    \includegraphics[width=.95\textwidth]{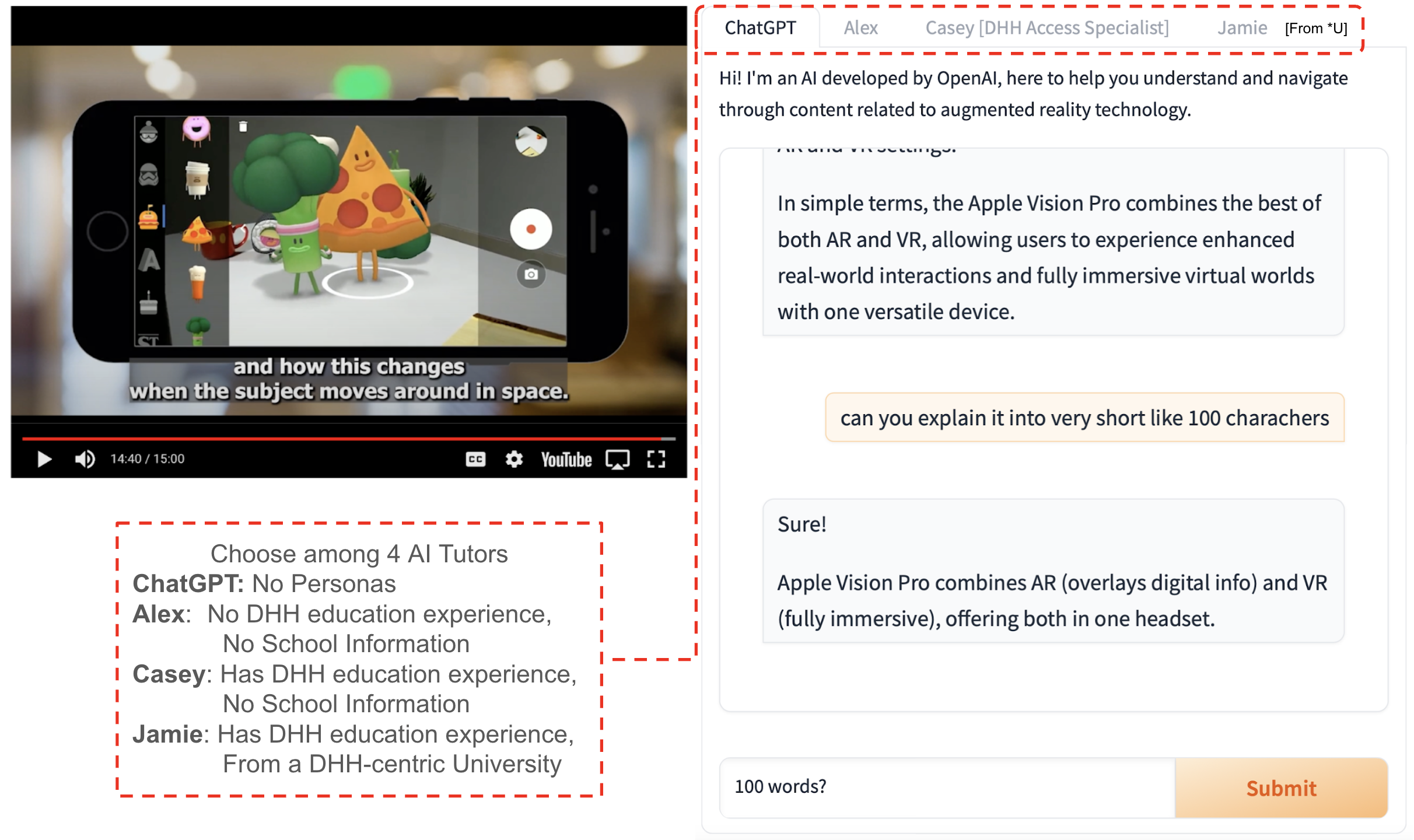}
    \caption{Prototype interface used for our user study. On the left side, there is a YouTube video player. On the right side, there is a chat window with four tabs, each representing one of the four AI tutors. Users may click on a tab to interact with the selected AI tutor. A one-sentence introduction of the selected AI tutor is shown above the chat window. This screenshot shows a sample interaction by P13, including their typo in their prompts. Note that the video player and chat window were resized on this screenshot for demonstration purposes, and the summarized differences among the four AI tutors were not available on the actual interface.}
    \label{fig:Interfacefig}
    \Description{This figure shows the interface of the prototyped online learning tool. On the left, there is a video player showing a screenshot from the 15-minute educational video about AR. The screenshot demonstrates two AR objects, a pizza and a broccoli, standing on a table, as seen on the screen of a smartphone. On the right of the video player, there is a chat window. On the top are four tabs: ChatGPT, Alex, Casey [DHH Access Specialist], Jamie [From *U]. The ChatGPT tab is chosen, with an introduction message of ChatGPT shown below it. Further below is the chat window that shows the chat messages sent by users with yellow background and messages from the AI tutor with gray background. On the bottom of the chat window, there is a text box with text "100 words" typed by a participant. Beside the text box is a button with text "Submit" on it. The tabs are marked red and connected to a text box explaining the difference across four AI tutors. ChatGPT has no personas; Alex has no DHH education experience and no school information; Casey has DHH education experience but no school information; Jamie has DHH education experience and is from a DHH-centric University.}
  \end{figure*}

Over the years, intelligent tutoring systems (ITS) using artificial intelligence (AI) technology have shown promise in education to create a personalized and adaptive learning experience \cite{carbonell1970ai} as well as supporting learners with diverse abilities \cite{michaud2000intelligent}. AI-based tutors \cite{graesser2001intelligent, rus2013recent}, for example, allow automated tutoring through conversation dialogue, where the AI serves as tutors. AI tutors were also developed to support DHH individuals learning written language (e.g., English) \cite{michaud2000intelligent} and sign language (e.g., ASL - American Sign Language) \cite{xu2007design, ritchings2012intelligent}.  

With the rapid growth of large language models (LLMs) in recent years, AI tutors saw new opportunities in providing dynamic interactive dialogue that combines the knowledge from LLMs and course materials for a more personalized learning experience \cite{schmucker2024ruffle, kakar2024jill}. LLMs also enabled incorporating personas, fictional characters that represent a specific group of people \cite{pruitt_2003personas}, to AI tutors, which shows promise in improving students' interaction experiences and educational performance \cite{ruan2019quizbot}.

However, there is a general lack of attention on how LLM-powered ITS may or may not address specific needs of DHH learners \cite{huffman2024we}, especially with the incorporation of personas. For example, LLM-powered personas have the potential of addressing the social needs of DHH learners by providing the support from the DHH community that they demanded \cite{lee2022struggle, gustafson2024survey, lynn2020review}. In addition, there is a lack of understanding on how DHH individuals engage with LLM-powered generative AI tools, and what accessibility concerns they may have toward such tools. While LLMs brought new opportunities to support DHH learners, it also brought risks such as biased training data and flawed AI models \cite{desai2024systemic}. 
% In addition, while there are works in designing personas for DHH individuals, to our knowledge, there is limited work on how LLM-powered persona may be incorporated in AI tutors to support DHH learners. 
% Additionally, there is limited understanding on the specific needs that may or may not be addressed by current LLMs.

To address the research gaps and understand the interaction between DHH learners and LLM-powered ITS in online learning, we evaluated a prototyped online learning tool, shown in Fig. \ref{fig:Interfacefig}, for DHH learners to interact with ChatGPT as well as three AI tutors with LLM-powered personas while watching online educational videos. The three AI tutors had different backgrounds in educating DHH learners: Jamie, an AI tutor who was from [Anonymized] university, a DHH-centric university; Casey, an AI tutor who is a DHH access specialist; and Alex, an AI tutor. Both Jamie and Casey had experience educating DHH students with different levels of knowledge toward DHH learners' preferences, whereas Alex had experience educating hearing students. To simulate the knowledge in DHH education experiences, we incorporated comments collected from DHH and hearing students and preferences of DHH learners into the prompt design of each AI tutor. Through a user study with 16 DHH learners, we answer the following research questions:

\begin{itemize}
    \item \textbf{RQ1}: How do DHH learners interpret and evaluate LLM-powered AI tutors with DHH education background as personas?

    \item \textbf{RQ2}: What interaction and accessibility preferences do DHH learners express when engaging with LLM-powered AI tutors? 

\end{itemize}

Our paper makes the following contribution to the HCI communities. 
1) We develop understanding in the perception and interaction between DHH learners and LLM-powered personas with different DHH education experiences. For example, DHH learners evaluated the community knowledge and position of the LLM-powered persona within the DHH community by asking questions relevant to their DHH identity. 2) We identify the diverse needs that were not addressed by current LLMs. For example, the request of simple and short responses in the persona prompts were perceived by some DHH learners as not detailed enough while still being too long for others. 3) We discuss the ethical and design implications for more inclusive LLMs. For example, we highlighted the importance of being cautious and considerate in designing ability-related background information in persona design within a diverse ability population that could shape their unique expectations.

\section{Related Works}

\subsection{AI Technologies for DHH Learners}

Over the years, researchers highlighted the need to adapt emerging technology, including AI technology, to support the educational experience for DHH learners \cite{alshawabkeh2023technology, wang2018accessibility, mack2023maintaining}. In the context of video-based online learning, prior studies have explored how AI technologies could be used to support the DHH learners to better take advantage of it. A common approach is the use of Automatic Speech Recognition (ASR) technology to support real-time captioning for DHH learners \cite{butler2019exploration, kushalnagar2012readability, lasecki2014helping}; however, using captions is not enough to address all needs that DHH learners have due to some DHH learners preferring the use of sign language \cite{berke2017deaf,chen2024comparison}. To this extent, there have been efforts in using AI for sign language capture, recognition, and translation \cite{papastratis2021artificial, bragg2019sign}. Collaborative efforts have produced large datasets for ASL recognition \cite{joze2018ms}, and advanced models like transformers have been applied to sign language translation \cite{camgoz2018neural, camgoz2020sign}. AI also plays a key role in providing personalized tools for learning English grammar \cite{cannon2011improving}, improving access in educational and work settings \cite{arroyo2024customization, benkerzaz2019study}, and generating questions for DHH students’ video-based learning \cite{huffman2024we}. More recently, studies also explored supporting DHH online learners with AI-based automatic emotion detection \cite{chen2024comparison} with a learning analytic tool. However, there is less attention in how AI could address the social needs for DHH learners, especially the support from teachers or fellow students within the DHH community \cite{lee2022struggle, lynn2020review}

% DHH students want support from same community.

Albeit the progress, potential risks remain in AI technology for DHH individuals. \citet{desai2024systemic} highlighted biases in sign language AI studies led by hearing, non-signing researchers, such as overemphasizing communication barriers, using non-representative datasets, and flawed AI models. Similarly, LLMs trained on hearing users' data may perpetuate discrimination and biases against minority groups such as DHH individuals \cite{mack2024they}. Prior studies also found DHH individuals are curious about AI technology \cite{coy2024inclusive} while having some fear due to unfamiliarity \cite{chen2024comparison}. A recent study \cite{huffman2024we} studied how individuals in DHH community used LLMs in their daily life, and they found that the use has been limited to language learning, mediating communication with hearing individuals, and simplifying information retrieval. In addition, DHH individuals with varying English proficiency may impact their perceived benefit and potentially lead to the risk of blindly trusting misinformation generated by the LLMs. However, a gap remains in understanding DHH individuals' accessibility and ethical concerns towards generative AI and LLMs. This study investigates how DHH users interact with text-based LLMs, i.e., ChatGPT \cite{gpt4}, assessing the design suggestions to address accessibility concerns.

\subsection{Intelligent Tutoring Systems for DHH Learners} 

One of the educational technologies powered by AI is intelligent tutoring systems (ITS), which could provide personal and adaptive learning experiences \cite{carbonell1970ai}. These AI-based tutors are designed to mimic the one-to-one experience between a student and a human tutor through personalized tutoring \cite{akyuz2020effects, shabrina2023investigating}, adaptive learning \cite{mousavinasab2021intelligent}, and conversational dialogue \cite{graesser2001intelligent, rus2013recent}. Over the years, ITS has been proven effective in education \cite{vanlehn2011relative, kulik2016effectiveness} and applied across various educational fields, such as language learning \cite{swartz2012intelligent} and science, technology, engineering, and mathematics (STEM) education \cite{graesser2018electronixtutor}. 
% Prior to the wide deployment of LLMs such as ChatGPT, generative AI has been used for content generation in ITS  \cite{di2005natural} such as generating formative feedback \cite{stamper2011enhancing, stamper2024enhancing}.  
The development of LLMs offered new possibilities in personalized tutoring and adaptive learning experience. For example, LLMs can generate quiz questions \cite{nguyen2024reference} personalized to a learner's understanding level \cite{maity2024generative}. LLMs also enabled AI-based tutors through interactive dialogue systems beyond traditional conversational agents (CAs). \citet{schmucker2024ruffle} developed Ruffle\&Riley, an AI tutor framework that allows educators to import learning materials and generate tutoring scripts through conversation. \citet{kakar2024jill} developed a virtual teaching assistant that leverages OpenAI's ChatGPT LLM \cite{gpt4} as well as instructor-approved learning materials, and they found that this approach maintained the benefits of CAs while suppressing some limitations of LLMs such as biases. In another study, LLM-based CAs also showed promise in fostering social interaction in online education \cite{kakar2024sami}, which could address the social needs from DHH learners \cite{lee2022struggle, lynn2020review, gustafson2024survey}.

Given the benefit of using ITS, prior research adapted ITS for DHH learners to learn languages. \citet{michaud2000intelligent} developed an ITS for DHH learners whose native language is ASL to help teach them written English. There are also ITS designed to teach sign language \cite{xu2007design, ritchings2012intelligent}. \citet{glasser2020accessibility} proposed designs for sign language-based CAs to support DHH users. However, there is still a general lack of addressing accessibility and specific needs for CAs \cite{divekar2021conversational}, such as the social support from the DHH communities. Additionally, there has been little work exploring how ITS could support DHH learners beyond learning language.

\subsection{LLM-powered Personas}

Personas are fictional characters created from data about a specific group of people \cite{pruitt_2003personas}. 
Personas typically include background information, goals, or behaviors, and are commonly used in product design, user experience research, and education \cite{pruitt_2003personas, goodwin_2001goal, oleson_2018pedagogical}. 
% Personas can help users focus their attention on a specific target audience, which can make users feel more empathy towards that group of people \cite{salminen_2020persona, pruitt_2003personas, miaskiewicz_2011personas}. 
% \citet{salminen_2022use} conducted a systematic literature review of persona and found 13 out of the 40 papers that met their inclusion criteria were focused on using personas to increase accessibility.
Recently, personas have been incorporated into LLMs \cite{salminen2024deus, ha2024clochat, sun2024building}. AI personas are artificial intelligence systems designed to embody specific characteristics, behaviors, and communication styles that align with particular user groups.
Given the growing body of evidence that suggests LLMs can simulate a range of human responses \cite{dubois2024_alpacafarm, park2023_generative}, designing personas to represent those responses likely extends our understanding of LLMs' abilities to follow specific guidelines.
% Prior research \cite{chiang2024_enhancing, li_2024user} used LLMs to provide feedback to participants from different perspectives enhancing their decision making.
We note that there are two areas of research focus regarding LLM-powered personas: one focuses on using LLMs to generate personas \cite{salminen2024deus, kocaballi_2023conversational, de2023writing, zhang2023personagen}, whereas the other area of research focuses on how personas act as CAs powered by LLMs \cite{markel2023gpteach, graesser2004autotutor, nye2014autotutor, ruan2019quizbot, ha2024clochat}. The latter use-case is how we use our LLM-powered personas in our study. A recent work \cite{ha2024clochat} explored how users customize the persona of an LLM-based CA through customized prompts, and they found that such customized personas could alter the responses from the LLMs as well as the perception of the users compared to LLMs without persona (e.g., ChatGPT).
\citet{ruan2019quizbot} suggests that applying personas to AI agents could improve students' interaction experience and educational performance.
% For example, \citet{ruan2019quizbot} found that when compared to a normal flashcard app, their tool QuizBot, helped students recall more information, improving their performance. Additionally, students reported feeling more engaged with QuizBot than the flashcard app \cite{ruan2019quizbot}. 

In educational settings, personas have been used to foster inclusivity in education by recognizing students from historically underrepresented communities, like students with diverse abilities \cite{salminen_2022use}, and examining how the traits of various student groups interact to influence learning experiences \cite{Brooks_2014, maier_2010using}. 
Prior research has also designed personas for DHH individuals. When designing personas for DHH participants, researchers have considered a variety of background information \cite{suzianti_2019implementation, nolte_2022creating} such as sign language ability, spoken language ability, technology proficiency \cite{nolte_2022creating}, and personality \cite{suzianti_2019implementation}. A recent work explored how users may benefit from interacting LLM-powered personas with different abilities and found such systems to help better understand the perspectives and build empathy towards individuals with different abilities \cite{sun2025persona}. However, to our knowledge, no research has considered how a persona having DHH teaching experiences as background information influences DHH participants interactions with the personas.

\section{Methods}

To answer the research questions, we conducted user studies with a prototype interface that allowed DHH learners to interact with multiple LLM-powered AI tutors with personas that had different experiences in DHH education. A total of 16 DHH participants were recruited. Participants were asked to watch a 15-minute educational video about Augmented Reality (AR) technology. They were able to interact with four AI tutors in individual chat windows, one was ChatGPT and the other three were AI tutors using role-playing ChatGPT. Two of the three AI tutors were given prompts that included various experiences in DHH education and response style. Participants completed a semi-structured interview session about their experiences and perceptions of the different AI tutors and how well they helped them learn. Our study was approved by our university institutional review board (IRB). Below, we present more details about this study.

\textbf{Positionality.} The research team consists of members with various backgrounds. The first author is a hearing individual with experience in research with the DHH community. He led through the process of designing the study material, including the prototype interface, conducting 12 of 16 studies in written and/or spoken English, and coordinated the data analysis with other authors. The second author is a hearing individual who has been learning American Sign Language (ASL) for two years and has experience in designing and conducting user studies with DHH and other ability-diverse participants. She helped with refining study material, observed half of the user studies, and discussed the findings. The third author is a hearing individual from a family with multiple Deaf members. He learned ASL from birth and is fluent in ASL. He joined the team after the study material was designed, conducted four user studies in ASL, and was involved in the data analysis process. This study was supervised by a hearing faculty member with experience in accessible HCI research.

\subsection{Prototype Design}

We designed a prototype interface for the user studies to allow participants to interact with four LLM-powered AI tutors with different experience in DHH education, all developed based on the GPT-4o model. 

\subsubsection{Online Learning Material}

We used a 15-minute educational video about AR technology as the online learning material for our study. We selected AR technology as the video topic because it is an emerging area that participants may have heard of, but are unlikely to be deeply familiar with in terms of technical details. This video was designed as a first lesson in a series of courses about AR technology. It covers history, application, and technical areas of AR technology. The video was well-perceived on Coursera\footnote{\url{https://www.coursera.org}} as an introductory level video on AR technology (4.5 out of 5, rated by over 3,800 learners). As a video was designed for general viewers, only spoken English was used by the presenters in the video. To make the video more accessible, we added error-free open captions to the video.

\subsubsection{Collecting Student Comments} \label{sec:cscw_comments}

Inspired by prior works on creating AI-based tutors using crowd data \cite{yang2021can} and suggestions to include real data in LLMs to support DHH learners \cite{huffman2024we}, we collected comments from DHH and hearing students on the 15-minute AR video prior to conducting the user study to simulate the educational experience for AI tutors on the selected online learning material. Specifically, we recruited 20 DHH learners (15 Deaf, 5 Hard-of-Hearing) from a DHH-centric university and 20 hearing learners from mainstream universities, both in the US. They were asked to watch the full video once and then reflect on their thoughts and feelings regarding any part of the video. They could watch as much of the video as they wanted to during the commenting process. Each comment contains a timestamp of the video and text comments. Note that three Deaf participants opted to make comments in ASL, which was later transcribed by an ASL interpreter into text. In total, 138 comments were made by DHH participants, whereas 78 comments were made by hearing participants.

\subsubsection{Persona Design} \label{sec:persona}

To answer our research questions, we designed LLM-based AI tutors with personas that varied in experience in educating DHH learners. These AI tutors were designed by three of the research team members based on prior findings on DHH learners' preferences in online learning \cite{chen2024motion, chen2024comparison, chen2024signmaku}. The prompt design was further revised during two pilot studies with a Deaf individual from a DHH-centric university who was majoring in accessible HCI and had research experience in interaction between DHH learners and AI technology. 
Note that our main goal was not to design a perfect AI tutor, but to offer different AI tutor design options for DHH learners to explore their preferences.
% Note that the main goal of the pilot studies was not to design a perfect AI tutor, but to understand the preferences from DHH learners that were missing in prior literature to offer different options for DHH learners to 

% To these extents, we used ChatGPT without personas as the baseline and designed three AI tutors with personas on different experience in educating DHH learners. 
% To answer our research questions, we designed three AI tutors with personas on different experience in educating DHH learners (with names generated by ChatGPT): \textbf{Alex}, \textbf{Casey [DHH Access Specialist]}, and \textbf{Jamie [From Gallaudet University]}. Additionally, we included ChatGPT as a fourth non-persona option to compare how DHH learners perceive AI tutors with or without personas in online learning. 
% The AI tutors were designed by three of the research team members based on prior findings on DHH learners' preferences in online learning \cite{chen2024motion, chen2024comparison, chen2024signmaku}. The prompts were further revised in two pilot studies with a Deaf individual from a DHH-centric university who was majoring in accessible HCI and had research experiences in interaction between DHH learners and AI technology. Note that the purpose of the pilot studies was not to design a perfect AI tutor, but rather to offer AI tutors designed for communities with different hearing ability to allow DHH learners explore their preferences when interacting with the LLMs. 

\begin{table}[]
\resizebox{\columnwidth}{!}{%
\begin{tabular}{l|lccc|}
\cline{2-5}
 &
  \multicolumn{4}{c|}{\textbf{AI Tutors}} \\ \hline
\multicolumn{1}{|l|}{Prompt Component} &
  \multicolumn{1}{c|}{\textbf{ChatGPT}} &
  \multicolumn{1}{c|}{\textbf{Alex}} &
  \multicolumn{1}{c|}{\textbf{\begin{tabular}[c]{@{}c@{}}Casey\\ {[}DHH Access \\ Specialist{]}\end{tabular}}} &
  \textbf{\begin{tabular}[c]{@{}c@{}}Jamie\\ {[}From a DHH-\\ centric University{]}\end{tabular}} \\ \hline
\multicolumn{1}{|l|}{\begin{tabular}[c]{@{}l@{}}Video Transcript \& \\ Description of images in video\end{tabular}} &
  \multicolumn{1}{c|}{Y} & 
  \multicolumn{1}{c|}{Y} & 
  \multicolumn{1}{c|}{Y} & 
  \multicolumn{1}{c|}{Y} \\ \hline
\multicolumn{1}{|l|}{Role} &
  \multicolumn{1}{l|}{} &
  \multicolumn{3}{c|}{\begin{tabular}[c]{@{}c@{}}Teaching Assistant in Information Technology, \\ Has Enthusiasm in AR technology\end{tabular}} \\ \hline
\multicolumn{1}{|l|}{\begin{tabular}[c]{@{}l@{}}Share students' thoughts \\ on the video\end{tabular}} &
  \multicolumn{1}{l|}{} &
  \multicolumn{1}{c|}{\begin{tabular}[c]{@{}c@{}}Hearing Students'\\ thoughts\end{tabular}} &
  \multicolumn{2}{c|}{DHH Students' thoughts} \\ \hline
\multicolumn{1}{|l|}{\begin{tabular}[c]{@{}l@{}}\textbf{Response Style:} \\ Share Personal Experience and \\ Explain concepts in a fun way\end{tabular}} &
  \multicolumn{2}{l|}{} &
  \multicolumn{1}{c|}{Y} &
  \multicolumn{1}{c|}{Y} \\ \hline
\multicolumn{1}{|l|}{\begin{tabular}[c]{@{}l@{}}\textbf{Response Style:} \\ Keep Language Short and Simple, \\ Ask questions related to the video\end{tabular}} &
  \multicolumn{3}{l|}{} &
  Y \\ \hline
\end{tabular}%
}
\caption{Prompt Designs for the four AI tutors used in our study. See Appendix \ref{appendix: prompt} for the exact prompt language used for each AI tutor.}
\label{tab:persona}
\Description{This table shows the prompt component of each AI tutor: ChatGPT, Alex, Casey [DHH Access Specialist], and Jamie [From Gallaudet University]. Video Transcript & Description of images in video are available in all four AI tutors. There are no other prompt component for ChatGPT. Role for Alex, Casey, and Jamie are the same: Teaching Assistant in Information Technology, Has enthusiasm in AR technology. Alex shares hearing students' thoughts on the video, whereas Casey and Jamie shares DHH students' thoughts. For response styles, Casey and Jamie would share personal experience and explain concepts in a fun way. Jamie would also keep language short and simple, and ask questions related to the video.}
\end{table}

Specifically, we designed three AI tutors with varying personas on experiences in educating DHH learners (with names generated by ChatGPT): \textbf{Alex}, \textbf{Casey [DHH Access Specialist]}, and \textbf{Jamie [From Gallaudet University]}. We also included ChatGPT (without a persona) as a fourth ``baseline'' AI tutor. Specifically, the AI tutors were designed to compare between having personas and not (ChatGPT vs. three AI tutors), between having experience in DHH education and not (Alex vs. Casey \& Jamie), and the specific background information regarding DHH education (Casey vs. Jamie). All four AI tutors were developed using GPT-4o \cite{gpt4}, which is widely known by the public and has the capability of interpreting videos. However, due to the technical limitations of GPT-4o model, we used audio transcripts and text-based summaries of images as input instead of the actual video. 
% the GPT-4o model could only use text-based audio transcripts as input rather than directly using the audio from the video, which had additional information such as pitch and tone of the narrators. As for the image input, we used GPT-4o model to generate a text summary of the images of the video into text, which is then used in each AI tutor's prompts. We decided to use text description rather than images as it would take too long for the GPT-4o model to interpret images from the full 15-minute video for each interaction. 
All four AI tutors (including ChatGPT) received the same transcripts and summary of image input as part of their prompts. Additionally, all four AI tutors were instructed to acknowledge that they were assisting DHH learners in learning AR technology from the video.

The three AI tutors with personas were designed as teaching assistants majoring in information technology with enthusiasm in AR technology. We did not specify the hearing ability of any personas, nor did we include avatars to any of the personas to avoid distracting learners from watching the video. The different prompt design of the AI tutors is summarized in Table \ref{tab:persona}, and the prompts used for each persona can be found in Appendix \ref{appendix: prompt}. Below, we describe the main components in the prompt design, and how the AI tutors differ from each other.

\begin{itemize}
    \item \textbf{Background information in DHH education experience.} A major difference across the AI tutors is how they described their experience in DHH education: Alex did not specify their experience in teaching DHH learners or affiliation with a school; Casey was shaped as a DHH access specialist without implying their school; Jamie was presented as being from Gallaudet University, a DHH-centric university. We differentiated this information to explore the preferences and perceptions toward AI tutors with different backgrounds.

    \item \textbf{Prior Students' Comments.} As described in Section \ref{sec:cscw_comments}, we collected hearing and DHH students' comments on the 15-minute AR video to simulate the education experience using real data. For Alex, we used hearing students' comments; for Casey and Jamie, we used DHH students' comments. The AI tutors were instructed to share previous students' thoughts about the video in their responses. These comments were used to better simulate the different DHH education experiences from different AI tutors.

    \item \textbf{Response Styles.} Prior studies suggested that the prompt design of personas could alter the response styles from the LLMs \cite{ha2024clochat}. To differentiate the response styles across the AI tutors, we added prompts to instruct how each AI tutor should format its responses, as summarized in Table \ref{tab:persona}. For Jamie, we used feedback of DHH students' preferences in a DHH-centric university based on prior studies \cite{chen2024motion, chen2024comparison, chen2024signmaku} as well as feedback from our pilot studies. For Casey, we selected the prompts that were designed based on observations from DHH students' comments. For Alex, we instructed it to keep responses similar to humans to differentiate from ChatGPT.
    
\end{itemize}

We compared the responses from each AI tutor on the same question during the pilot studies to evaluate whether each AI tutor reflected their prompts. Some sample responses on the same questions, as collected during our user studies, can be found in Appendix \ref{appendix:sampleresponse}.

\subsubsection{Interface Design} \label{sec:interface}

The prototype interface, shown in Fig. \ref{fig:Interfacefig}, was built using Gradio\footnote{\url{https://www.gradio.app}} package in Python. We used a side-by-side design with a YouTube\footnote{\url{https://www.youtube.com}} video player on one side and the chat window on the other side. A similar side-by-side design with video player on one side and text-based content on the other side was found in other online video platforms such as YouTube and Twitch, as well as in online educational tools such as ClassTranscribe \cite{angrave2020benefits}. Within the chat window, users may choose the AI tutor they would like to interact with by selecting the corresponding tab on the top.
The following introduction, as pre-generated by each AI tutor, is available above the chat window when the corresponding tab is selected:

\begin{itemize}

    \item \textbf{ChatGPT}: \textit{Hi! I'm an AI developed by OpenAI, here to help you understand and navigate through content related to augmented reality technology.}

    \item \textbf{Alex}: \textit{Hi, I'm Alex, an enthusiastic teaching assistant specializing in Information Technology, particularly passionate about Augmented Reality (AR) technology.}

    \item \textbf{Casey [DHH Access Specialist]}: \textit{Hi, I'm Casey, a teaching assistant with a passion for Information Technology and Augmented Reality, experienced in supporting Deaf and Hard-of-Hearing students.}

    \item \textbf{Jamie [from Gallaudet University]}: \textit{Hi, I'm Jamie, a teaching assistant at Gallaudet University specializing in Information Technology, and I'm passionate about Augmented Reality (AR) technology.}

\end{itemize}

When a user submits a message, the persona prompts (as system prompt) together with the chat history will be submitted to the GPT-4o model, and the response received from the model will be subsequently printed on the chat window. The interaction with each AI tutor is independent of the others, meaning that one AI tutor will not know the chat history or prompt from another AI tutor. Note that users could not see the persona prompts for any AI tutor, nor could they alter these prompts during interaction.

\subsection{User Study}

\begin{figure*}[t]
    \includegraphics[width=.9\textwidth]{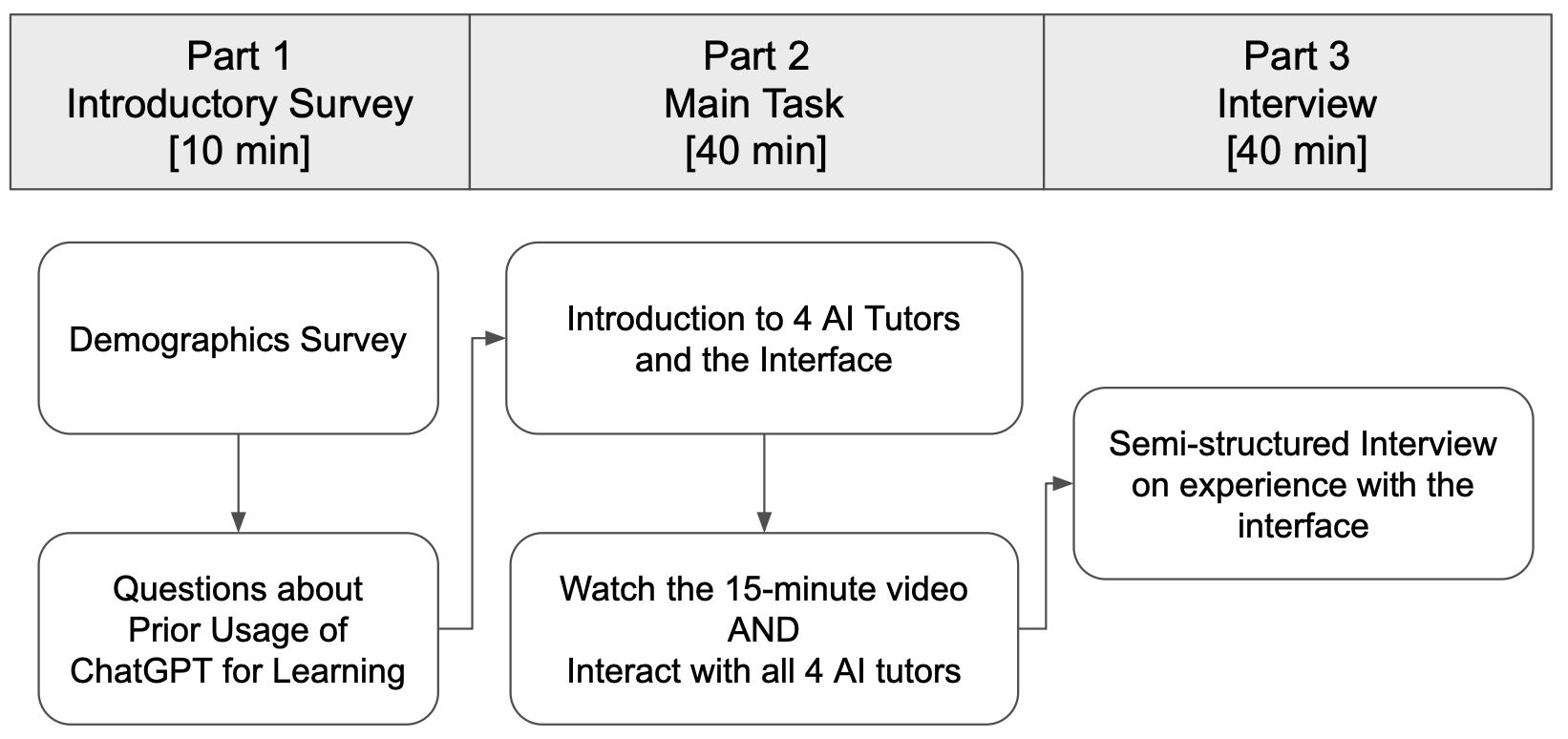}
    \caption{Study Procedure.}
    \label{fig:Processfig}
    \Description{This is a flow chart that shows the study procedure. On the top are four rectangles representing three parts of the study: Part 1 Introductory Survey [10 min], Part 2 Main Task [40 min], Part 3 Interview [40 min]. Under Part 1, there are two components. The first one is "Demographics Survey." An arrow points from this component to the second component under Part 1: "Questions about Prior Usage of ChatGPT for Learning." An arrow points from this component to the first of two components under Part 2: "Introduction to 4 AI Tutors and the Interface." An arrow points from this component to the second component in Part 2: "Watch the 15-minute video AND Interact with all 4 AI tutors." An arrow points from this component to the only component under Part 3: "Semi-structured Interview on Experience with the Interface."}
  \end{figure*}

To evaluate our prototype interface, we conducted user studies with DHH learners. Each study session lasted between 1 and 2 hours, with most studies completed within 1.5 hours. We compensated participants at a rate of \$30 per hour. The study procedure is shown in Fig. \ref{fig:Processfig}.

\subsubsection{Participants} 

A total of 16 DHH participants (referred to as P1 to P16 in the remainder of this paper) were recruited from a mainstream university (13 participants) and a DHH-centric university (3 participants), both in the US. Participants in the mainstream university were recruited through an email list of current DHH students. Participants in the DHH-centric university were recruited through word-of-mouth. To qualify for the study, the participant must be 1) at least 18 years old; 2) be a current or recent (graduated in the last 6 months) college student, OR currently engaging in online learning courses.

Out of the 16 participants, 5 self-identified as males and 11 as females. Their age was between 18 and 37. Six were d/Deaf, nine were Hard-of-Hearing, and one self-identified as hearing. The hearing participant was born as deaf and reported that they had cochlear implants soon after birth. Four participants learned ASL in preschool, seven had learned ASL in middle school, high school, and/or college, and five participants never learned ASL. Two participants also learned sign language other than ASL. Three participants had at least one immediate family member who was DHH. Nine of the participants were undergraduates, four were master's students, and three were doctoral students. As for ethnicity, six were White, four were Latino/Hispanic, three were Asian, two were Indian, and one was African American. Participants had a wide range of majors, including computer engineering, information technology, English, neuroscience, health technology, education, landscape architecture, biochemistry, and physics. As for experiences using LLMs, two participants (P3 and P5) reported that they never used any LLMs (including but not limited to ChatGPT), while seven participants reported using LLMs for learning at least once per week. A detailed description of the participants' demographics can be found in Appendix \ref{appendix:demo}.

\subsubsection{Study Procedure}

All studies were conducted via Zoom, an online communication platform. The study material was presented in written English. To accommodate our participants with different hearing ability and language preferences, we gave participants options to choose their most comfortable language for communication with the researchers during the study sessions. Four participants completed the study using ASL and were conducted by the team member fluent in ASL. The other participants communicated in spoken English (with auto-caption) and/or written English with a hearing researcher.

The study procedure is shown in Fig. \ref{fig:Processfig}. Upon greeting the researchers, participants first completed the consent form and were informed that the study sessions would be recorded but anonymized afterwards. Participants were also asked to share their screen during the study for researchers to observe their interaction using the prototype. 

\textbf{Part 1: Introductory Survey.} After the consent process, participants completed an introductory survey, which included demographic questions and questions about their prior usage of ChatGPT or other generative AI services. Two participants mentioned that they never used any generative AI services. Four participants also mentioned that they have never heard about Gallaudet University, which the researcher explained to them that Gallaudet University is a DHH-centric university.

\textbf{Part 2: Main Task.} Participants were introduced with the interface and the four AI tutors that they could interact with. We described the AI tutors with personas as ``role-playing ChatGPT experiences.'' We provided a one-sentence introduction for each AI tutor that was consistent across user studies (described in Section \ref{sec:interface}). Then, they were given the main task: watch a 15-minute educational video about AR and make at least one interaction with each AI tutor. An interaction is one prompt that was sent to an AI tutor. Participants were told that they could interact with each and every AI tutors at any time during the task and as many times as they wanted. We did not randomize the AI tutors, but presented them in the order of least to most amount of persona prompt as shown in Table \ref{tab:persona}; however, participants were not required to interact in the same order as presented. We gave them a 40-minute time limit to allow adequate time to watch the video and interact with all AI tutors while controlling the overall time for the study sessions. Once participants notified the researcher they were done interacting with the AI tutors, the interviewer moved onto the next part of the study. Note that one participant (P15) did not interact with Alex proactively, and we did not force them to interact.

\textbf{Part 3: Interview.} Participants completed a semi-structured interview session to discuss their experience with the prototype interface. Sample questions include: \textit{Do you have any suggestions to improve your chat experience with [AI Tutor name]}; \textit{did you have different expectations from different tabs before interacting with them}; and \textit{did the background information impact how you interact with different tabs}. To help participants answer the questions in their most comfortable language modality, participants may choose to complete the interview questions using a Google Form, or answer questions asked in ASL or spoken English (with auto caption). They may use ASL, written English, spoken English, or any combination of them to answer the questions. We also invited participants to make further interaction with the AI tutors if they wish to.

\subsubsection{Data Analysis}

We used qualitative approach to analyze the data and answer our research questions.

\textbf{Interview.} The interview responses were used to understand the reasons behind their perception toward AI tutors (RQ1) as well as their design suggestions (RQ2). The responses in spoken English were automatically transcribed by Zoom during the study, whereas the responses in ASL were transcribed by a researcher. We then conducted an inductive thematic analysis \cite{braun2006using} to analyze the transcripts. Specifically, two researchers open-coded one interview independently, discussed the results, and developed an initial codebook. The remaining transcripts were analyzed by one of the two researchers, and the research team frequently discussed and revised the codebook accordingly. Sample codes include: \textit{trustworthiness toward one AI tutor over others}, \textit{improving responses from AI tutors}, and \textit{specific AI tutor background information found useful}.

\textbf{Interaction logs.} Overall, participants sent 182 interactions (Mean=11.4, SD=6.12) during the main task, while 11 participants also made additional interactions during interviews to further evaluate the responses from the AI Tutors. An interaction is considered as a prompt sent to one of the AI tutors that received responses. Regarding each AI tutor, participants made 49 interactions with ChatGPT (Mean=3.1, SD=3.09), 36 interactions with Alex (Mean=2.3, SD=1.34), 42 interactions with Casey [DHH access specialist] (Mean=2.6, SD=1.96), and 54 interactions with Jamie [From Gallaudet University] (Mean=3.4, SD=2.03). 

We conducted thematic analysis on the content of the questions asked by the participants during the main task with a focus on whether each question is related to the video content and DHH community. Since some of the interactions during the interviews are based on instructions from the researchers, we only analyzed the interactions during the main task which represented their natural behaviors. Two researchers independently coded all questions. The Cohen's Kappa suggested good agreement (>80\%) on the coding results, and the two researchers then discussed on the remaining questions to reach agreements.

% We analyzed the questions asked by the participants during the task

% Since the focus of our research questions is not about the interaction between DHH learners and AI tutors, we did not conduct formal analysis on the interaction logs, but rather used them as examples to contextualize the interview findings. Below, we describe the basics of the interactions across four AI tutors. \haocong{Will revise this section by adding thematic analysis to the chat questions.}

\section{Findings}

% Note: Jamie is Gallaudet teacher; Casey is DHH teacher; Alex is Hearing teacher.

\subsection{RQ1: Interpreting and Evaluating LLM-powered personas} \label{sec:rq1}

% \subsubsection{Casey and Jamie were expected}

\subsubsection{Encouraging DHH-related questions}

% \haocong{New findings from chat log analysis.}

Out of the 182 interactions between participants and all four AI tutors during the main task, 45 interactions (24.7 \%) from 10 participants were related to questions about accessibility and DHH communities. Specifically, 16 interactions were with Casey, 16 with Jamie, eight with ChatGPT, and five with Alex. These questions include how AR technology can benefit DHH individuals, e.g., \textit{how is augmented reality benefitting the deaf people. In what ways is it giving an immersive experience to them}; how to use AR to help learning ASL, e.g., \textit{How can someone use AR when learning STEM signs and concepts in ASL?}; and how the AI tutors could support DHH learners, e.g., \textit{Rather than caption and subtitles, what could you do to provide assistance when I (as a deaf) watch the video}.

% 8 5 16 16

% By further looking into the interaction patterns of these questions, these questions were 
Participants mentioned during the interview that the background information (having DHH education experience or not) impacted the questions they would ask to each AI tutor, e.g., ``\textit{When I had a question regarding hearing, I asked Casey who is the DHH Access specialist}'' (P09). P03 asked DHH-related questions to Casey only, e.g., \textit{how AR can help in DHH community}, and asked non-DHH-related questions to Alex and ChatGPT, e.g., \textit{why are you passionate about AR and what is your goal of using AR}. They explained their strategy during the interview, ``\textit{I would ask Casey and Jamie because they know more about DHH. I think ChatGPT and Alex might not know because their focus area might answer differently.}''

Participants also altered the order of interaction with the AI tutors in order to ask DHH-related questions to Casey or Jamie. Since all participants received the AI Tutors with the same order (as shown in Figure. \ref{fig:Interfacefig}), it is expected that participants would interact with them in this order. However, six participants skipped Alex and/or ChatGPT to interact with Casey or Jamie first. For example, P01 first interacted with Jamie with the question \textit{How can augmented reality help accommodate those who are hard-of-hearing or deaf}. They then went back to ChatGPT to ask, \textit{How can augmented reality contribute to everyday life}. During the interview, they explained their different expectations to different AI tutors based on their background information, ``\textit{I'm assuming Jamie and Casey were more implemented with like deaf and hard-of-hearing people. ... You would have more perspective on, you know, that community and provide it in the prompt.}''

In fact, one participant, P15, who was discussing with the AI tutors about using AR technology for ASL learning, skipped Alex and interacted with Casey and Jamie after ChatGPT. They explained their reasons during the interview, ``\textit{for Alex, based on the background, because it didn’t say anything about ASL, I evaluated it, and I thought it probably couldn’t answer my questions, so I just skipped it. ...  it wasn’t worth my time. I just thought the last two personas [Casey and Jamie] would have better answers, so I decided to use them instead}.'' 
% Meanwhile, P15 initially doubted whether Casey could answer her questions, but ``\textit{trust what the third persona [Casey] says}'' after some interactions. 

% Within these questions, participants asked more questions about accessibility topics, including topics related to DHH community, to Casey and Jamie, two AI tutors with DHH education experiences. [Will add numbers here.] These questions include how AR technology can benefit DHH individuals.
% Participants asked more questions about DHH community to Casey and Jamie.

\subsubsection{Evaluating knowledge about the DHH community}

% \haocong{Existing findings from old RQ1, will revise language.}

The background information across different AI tutors represented their different knowledge level and position in the DHH community.
During the interview, participants discussed their perceptions toward this background information and how their expectations to their responses were impacted. 
Specifically, participants reported that they expected Casey and Jamie to have more knowledge about questions on DHH communities. 
For example, P01 asked all four AI tutors the same question, \textit{How can augmented reality help accommodate those who are hard-of-hearing or deaf}. They found the responses from four AI tutors to be slightly different from each other, and Casey provided the best response tailored to DHH individuals, ``\textit{the explanation Casey provided for my prompt provided extra detail such as including feedback regarding how DHH people felt about augmented reality being implemented for accommodations.}''

Meanwhile, three participants mentioned that Casey and/or Jamie failed to meet their expectations on their knowledge with the DHH communities. P8, for example, asked Casey the question \textit{How can AR improve IEP [Individualized Education Program] and accommodated learning plans for students in elementary school?} While Casey acknowledged the question, it ``\textit{failed to actually give me any useful information (specifically about DHH peoples) about the question}.'' P8 further elaborated, ``\textit{I think considering that it is a DHH specialist, it should frame everything in a context that could apply to DHH in some way, if that makes sense. That, or I'm not fully understanding what the point of this AI is.}'' P13, on the other hand, was not satisfied with the language style used by Jamie that claimed to be from Gallaudet University, ``\textit{if they were from Gallaudet University, people tend to explain very concise and they summarize a lot, and I would expect the responses from Jamie to be short. But they were really long, and people from Gallaudet University typically don’t prefer to read really long messages}.'' 

% Participants were still not fully satisfied with Casey and Jamie regarding knowledge on DHH community.

% Participants expect Jamie to have behaviors similar to someone from GU, but were not satisfied.

\subsubsection{Assessing the position within the DHH community}

% Only mentioning from GU does not tell their exact position within the DHH community. Participants would like more details such as ASL skills and how long they have been in the community.

% Participants from GU questioned the background of Jamie.

% \haocong{Existing findings from old RQ2}

Participants used the AI tutors' background information and their responses to assess their position within the DHH community. The description of Jamie from a DHH-centric university, in particular, led participants who were from the same university to guess Jamie's detailed identity. For example, P13 was suspicious that the content mentioned by Jamie did not match their impression with people from Gallaudet University, ``\textit{this person‘s supposed to be from Gallaudet University and I’ve never heard anyone from Gallaudet University mention that AR tool, so I was a little suspicious}.'' These expectations would impact how they interact with the AI tutors. P12 considered Jamie as a professor rather than a fellow student like Alex, and would be more hesitant about what to discuss with Jamie: ``\textit{I assume  Jamie's background, you know more professional, more like a teacher, more academic ... The fact that Jamie is from Gallaudet University itself means that it could be someone that I’ve talked to before ... it can’t be a conversation with someone else, like I wouldn’t have a conversation about my life with Jamie or anything that wasn’t academic}.'' They were also more concerned that their conversation with Jamie could be seen by someone from the university, ``\textit{if Jamie's from Gallaudet University, I’m wondering, can other teaching assistance or can the institution of Gallaudet University like see my questions? ... that would make me may be a little more hesitant to talk to Jamie because I wouldn’t be comfortable if Gallaudet University was watching me interact with this}.''

Participants suggested more transparency about how these AI tutors are situated within the DHH community rather than only mentioning their experience educating DHH learners. For example, P15 found the descriptions of Casey being a DHH access specialist and Jamie from a DHH-centric school not enough to fully understand their position within the community, as they commented about Jamie, ``\textit{For Jamie I understand that Gallaudet University is a big part of the deaf community and I feel like Jamie, and you know Gallaudet University clearly accepts Jamie which means that something is there, but we can’t just assume all the time we still need to know a little bit more background like their ASL skills and how long they’ve been a part of the deaf community}.''

It is worth noting that none of the participants mentioned the demand to see how the AI tutors were developed or tuned, but only relied on the description of their personas to form their impressions.

\subsection{RQ2: Accessibility preferences in LLM-powered AI tutors} \label{sec:rq2}

% \subsubsection{LLMs should support the multimodality of sign language}

\subsubsection{Supporting the Multimodality of Sign Language}

% \haocong{Existing findings from old RQ2}

\begin{figure*}[t]
    \includegraphics[width=.95\textwidth]{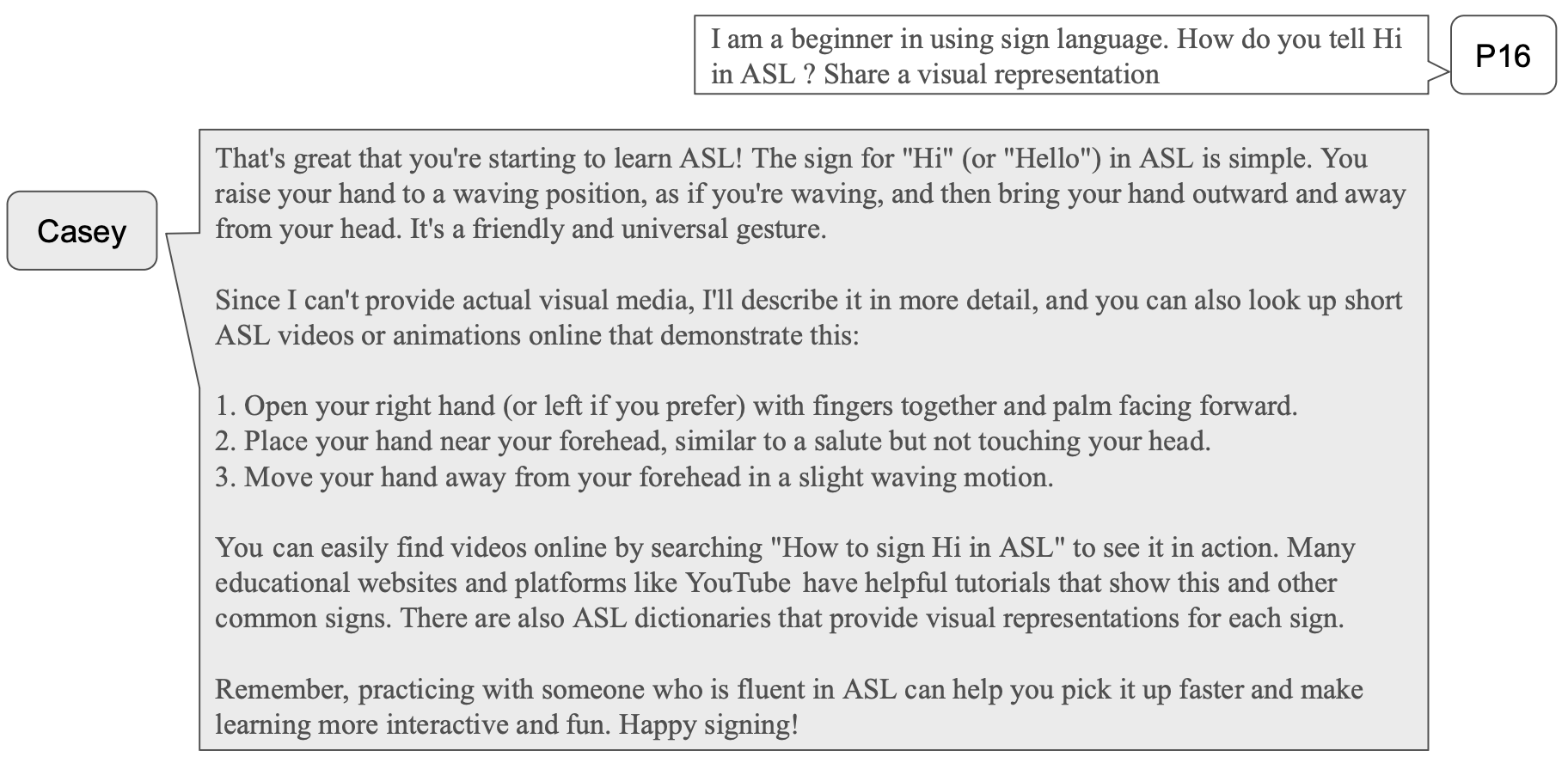}
    \caption{Part of interaction between P16 and Casey, where P16 asked Casey to demonstrate ASL through visual representation. Casey acknowledged that it was unable to produce visual output and described how to sign in English.}
    \label{fig:asl_p16}
    \Description{This figure shows chat interaction between P16 and Casey. P16 asked, "I am a beginner in using sign language. How do you tell Hi in ASL ? Share a visual representation." Casey responded, "That's great that you're starting to learn ASL! The sign for "Hi" (or "Hello") in ASL is simple. You raise your hand to a waving position, as if you're waving, and then bring your hand outward and away from your head. It's a friendly and universal gesture. Since I can't provide actual visual media, I'll describe it in more detail, and you can also look up short ASL videos or animations online that demonstrate this: 1. Open your right hand (or left if you prefer) with fingers together and palm facing forward. 2. Place your hand near your forehead, similar to a salute but not touching your head. 3. Move your hand away from your forehead in a slight waving motion. You can easily find videos online by searching "How to sign Hi in ASL" to see it in action. Many educational websites and platforms like YouTube have helpful tutorials that show this and other common signs. There are also ASL dictionaries that provide visual representations for each sign. Remember, practicing with someone who is fluent in ASL can help you pick it up faster and make learning more interactive and fun. Happy signing!"}
  \end{figure*}

During the interaction, three participants (P13, P14, P16) requested the AI tutors to explain a concept in ASL. Because the GPT-4o model cannot generate images or videos, a description of how to sign certain words in ASL was generated instead. Participants found these responses underwhelming without visual demonstration. P16, who was a beginner in ASL, asked Casey to teach them to say ``hi'' in ASL during the interview, as shown in Fig. \ref{fig:asl_p16}. In return, Casey mentioned that it could not provide actual visual media but described how to sign in text and suggested looking for videos demonstrating the sign. P16 found the lack of visual demonstration in the response and wished that could be implemented, ``\textit{since I don't know sign language, and I want to actually learn it ... It would be helpful to have a content that is translating text to a graphical or sign language approach.}'' P12, who was fluent in ASL, hoped the AI tutors could provide answer in ASL to easily demonstrate to their friends, ``\textit{It’d be nice if ChatGPT would provide an answer in ASL, because sometimes I use [ChatGPT], it’s great. But my friends will ask me what did it say and maybe they don’t understand, and so I have to sign it so they can understand it better.}''

Participants also wanted to be able to use ASL to ask questions of the AI tutors, as it would allow them to better express themselves. P12 said, ``\textit{because sometimes I want to ask someone but I’m not sure how to phrase it in English and it is a little awkward, whereas in ASL I’m ready, and I can ask the question immediately. So, if you have the option to let me ask the question ASL it would be a lot more accessible.}''

% Participants demand sign language (visual) as input and output.

% Participants want LLMs to use ASL gloss in their responses.

\subsubsection{LLM Response Style: More Details vs. Simple and Short}

% \textbf{More details vs. Simple and short}

% \haocong{Existing findings from old RQ2, will need to revise language.}

In Jamie's prompt, the AI tutor was instructed to keep its response short and easy to understand based on feedback from the pilot study. However, during our user studies, we observed mixed preferences between more detailed and more concise responses from AI tutors.

% \textbf{Responses with More Details.}
Four participants mentioned that they preferred the more detailed responses from AI tutors other than Jamie, as Jamie's responses were typically shorter and less detailed. 
% P1 said, ``\textit{I did use a similar prompt for the AI tools I was provided, and I have found that the answer Jamie provided to me wasn't as detailed compared to the other tools.}'' 
P2 mentioned, \textit{I would prefer it to be more detailed rather than it to be more vague... because it gives me a lot more information, and then I can decide what part I'd ask more questions about}.'' Participants also mentioned that the outline style that ChatGPT used in its response made it easier to locate the information they were looking for. P1 said, ``\textit{Additionally, AI tools such as ChatGPT and Casey provided headings for each section their answer provided me unlike Jamie, which made their answers a little easier to read.}''

\begin{figure*}[t]
    \includegraphics[width=.95\textwidth]{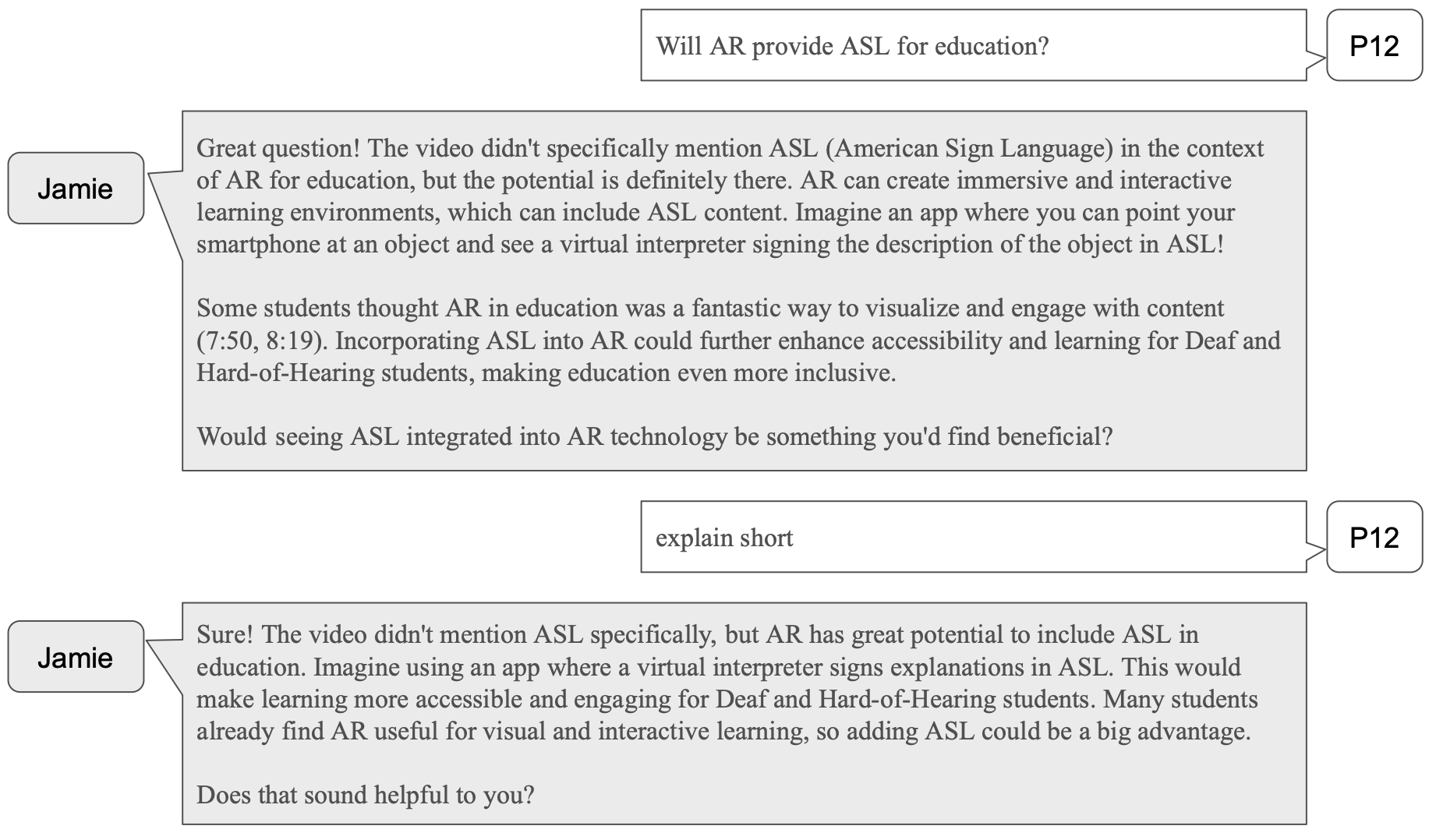}
    \caption{Part of interaction between P12 and Jamie. Although Jamie’s response was relatively short, P12 still asked Jamie to shorten it further.}
    \label{fig:short_p12}
    \Description{This figure shows chat interaction between P12 and Jamie. P12 asked, "Will AR provide ASL for education?" Jamie responded, "Great question! The video didn't specifically mention ASL (American Sign Language) in the context of AR for education, but the potential is definitely there. AR can create immersive and interactive learning environments, which can include ASL content. Imagine an app where you can point your smartphone at an object and see a virtual interpreter signing the description of the object in ASL! Some students thought AR in education was a fantastic way to visualize and engage with content (7:50, 8:19). Incorporating ASL into AR could further enhance accessibility and learning for Deaf and Hard-of-Hearing students, making education even more inclusive. Would seeing ASL integrated into AR technology be something you'd find beneficial?" P12 then asked, "explain short." Jamie responded, "Sure! The video didn't mention ASL specifically, but AR has great potential to include ASL in education. Imagine using an app where a virtual interpreter signs explanations in ASL. This would make learning more accessible and engaging for Deaf and Hard-of-Hearing students. Many students already find AR useful for visual and interactive learning, so adding ASL could be a big advantage. Does that sound helpful to you?"}
  \end{figure*}

\begin{figure*}[t]
    \includegraphics[width=.95\textwidth]{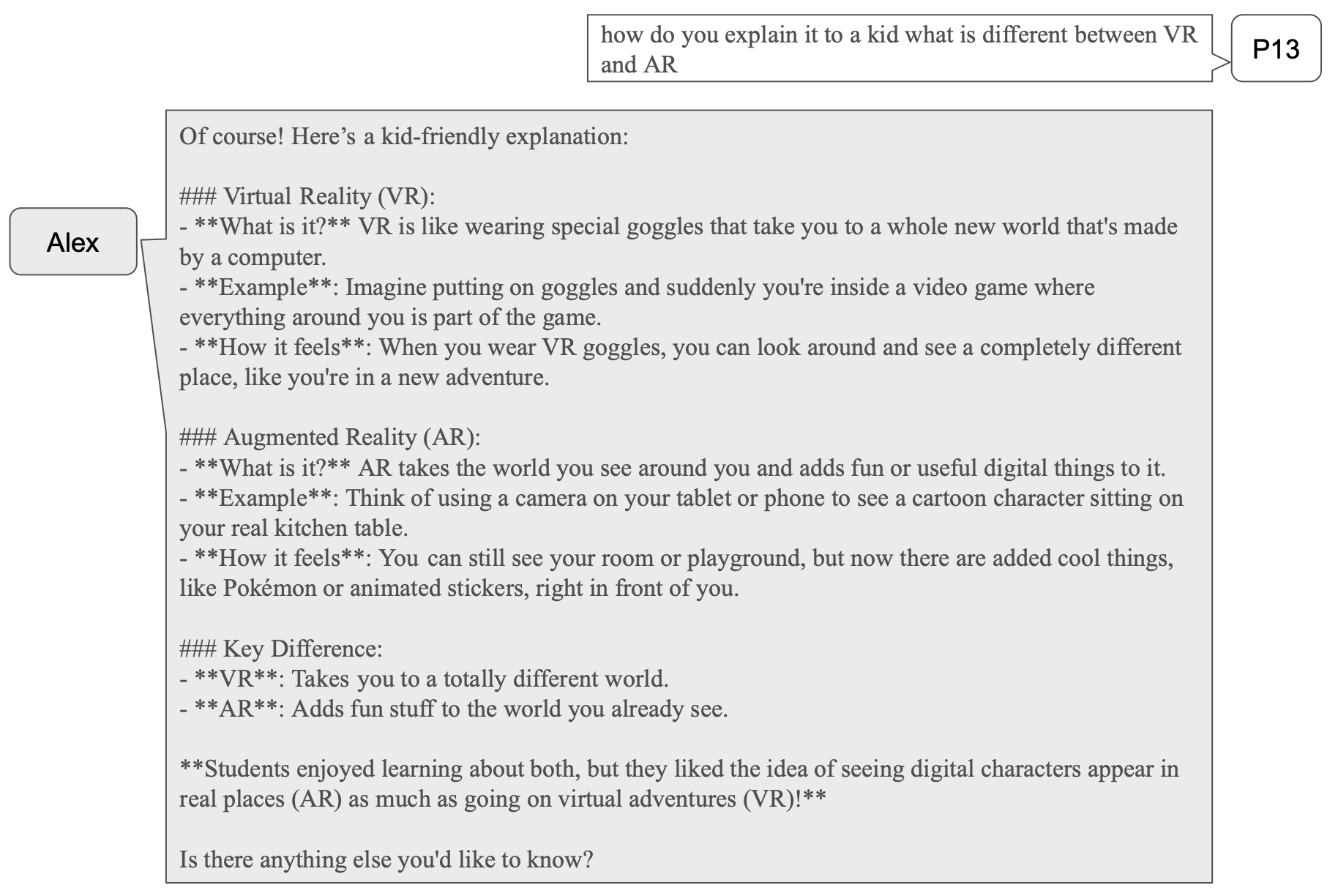}
    \caption{Part of an interaction between P13 and Alex, in which P13 asked Alex to explain the difference between AR and VR to a kid.}
    \label{fig:kid_p13}
    \Description{This figure shows chat interaction between P13 and Alex. P13 asked, "how do you explain it to a kid what is different between VR and AR." Alex responded, "Of course! Here’s a kid-friendly explanation: ### Virtual Reality (VR): - **What is it?** VR is like wearing special goggles that take you to a whole new world that's made by a computer. - **Example**: Imagine putting on goggles and suddenly you're inside a video game where everything around you is part of the game. - **How it feels**: When you wear VR goggles, you can look around and see a completely different place, like you're in a new adventure. ### Augmented Reality (AR): - **What is it?** AR takes the world you see around you and adds fun or useful digital things to it. - **Example**: Think of using a camera on your tablet or phone to see a cartoon character sitting on your real kitchen table. - **How it feels**: You can still see your room or playground, but now there are added cool things, like Pokémon or animated stickers, right in front of you. ### Key Difference: - **VR**: Takes you to a totally different world. - **AR**: Adds fun stuff to the world you already see. **Students enjoyed learning about both, but they liked the idea of seeing digital characters appear in real places (AR) as much as going on virtual adventures (VR)!** Is there anything else you'd like to know?"}
  \end{figure*}

% \textbf{Simple and Short Responses.}
Meanwhile, some participants were disappointed that the responses from Casey and Jamie were longer than they expected despite being described as having experience in DHH education.
% Given that Casey and Jamie described themselves as having experience in DHH education, some participants expected their responses to be shorter and more concise. However, they were disappointed that these responses were longer than they expected. 
For example, P4 shared their frustration, ``\textit{they all had really long responses, and I really thought that Casey and Jamie would provide short responses, but that wasn’t necessarily the case, and I thought Jamie would be the same as ASL gloss\footnote{ASL gloss is written approximation of ASL that uses English words to label signs and provide grammatical notes.}, which just means short and concise, and I thought Casey could be short, but may be more connected to the subject of AR, but they seemed pretty similar, and there were long blocks of text for their response}.''
% P13 further explained, ``\textit{if they were from Gallaudet University, people tend to explain very concise and they summarize a lot, and I would expect the responses from Jamie to be short. But they were really long, and people from Gallaudet University typically don’t prefer to read really long messages}.'' 

In order to tune the responses, five participants requested the responses from AI tutors to be simplified or shortened using prompts such as \textit{can you explain short?} (P12) or \textit{explain a gyroscope in simpler terms please} (P7). Fig. \ref{fig:short_p12} shows an example of P12 requesting Jamie to shorten its responses.
P13 used words ``short,'' ``simple,'' or specific word limit (e.g., ``100 words'') eight times across their 44 interactions across all four AI tutors to shorten their responses. They shared their frustration during the interview, ``\textit{when I said short, it still gave me a long answer. It wasn’t summarized so I need to explain specifically I want you to explain in 100 characters or 50 words. So I had to be very specific}.''
We also observed two participants, P12 and P13, asking ChatGPT and Alex to explain certain AR concepts mentioned in the video to a kid. A sample interaction between P13 and Alex is shown in Fig. \ref{fig:kid_p13}. P12 explained their rationale writing prompts in this way, ``\textit{the reason is sometimes well I don’t know a lot about AR and sometimes if I don’t understand the English. I’ll ask it to respond like it was giving an answer to a child. That way I can understand it a bit better, because when they explain when they provide an answer to a kid, they’re usually more summarized, and it is more understandable to me so sometimes that’s what I ask it to do because I found that works best}.''

P4, who had attended both mainstream school and DHH-centric school, summarized their struggle to find the right balance between detailed and concise responses: ``\textit{Personally, I went to an immersion school, so I’m used to talking to hearing people. And Gallaudet University was the first time I was in a deaf environment. Sometimes I get a little frustrated because the English education level is sometimes a little different, so it’s hard to find the right balance because sometimes I want more concrete examples or more information}.''

% In Jamie's prompt, we instructed it to keep responses simple and short following feedback from pilot studies.

% Some participants found Jamie's responses to be not detailed enough to be useful.

% Other participants found Jamie's responses to be still too long, and developed their own strategies to make it even shorter.

\subsubsection{LLM Response Style: Personal Support vs. Broader View}

% \textbf{Providing personal feedback.}
Participants mentioned that the positive support and personal feedback from the AI tutors, particularly Casey, made them feel more willing to continue the conversation. P14, who reported frequently using ChatGPT for learning, preferred asking questions to Casey over Alex and Jamie because the response felt warmer and more supportive: ``\textit{loved it [Casey] in attitude and answering the Qs! She is warm (greeting, name-calling), the answers she provided sometimes show in for people with disability (show text, visual part and accurate paragraph separating), and somehow she could provide more details for helping understanding}.'' P9 also liked Casey providing personal feedback in their responses, ``\textit{I liked interacting with Casey the most. The way it broke down my question in order to address every single thing as well as provide previous thoughts, personal feedback, and even gave me more helpful information that I had asked for was amazing}.''

% \textbf{Less personal responses leading to broader views.}
There were a few participants who preferred ChatGPT over the other AI tutors due to its direct, impersonal tone. P11 found ChatGPT easiest to gather information from, ``\textit{For some reason I liked the dispassionate nature of ChatGPT a bit better, because I was more interested in getting information than "talking to a person}.'' Similarly, P14, who liked Casey's personal feedback, also found ChatGPT's response helpful for them: ``\textit{Compared to Casey, ChatGPT is less warm, such as did not call my name and greet me, but I love the conclusion of the answer it gave which others did not ... ChatGPT could give me broaden and details answer}.''

% Some participants liked Casey providing a personal touch.

% Other participants said that less personal responses could lead to boarder view.

\subsubsection{Eliciting thinking by asking follow-up questions}

% \textbf{Asking more relevant questions.} 

\begin{figure*}[t]
    \includegraphics[width=.95\textwidth]{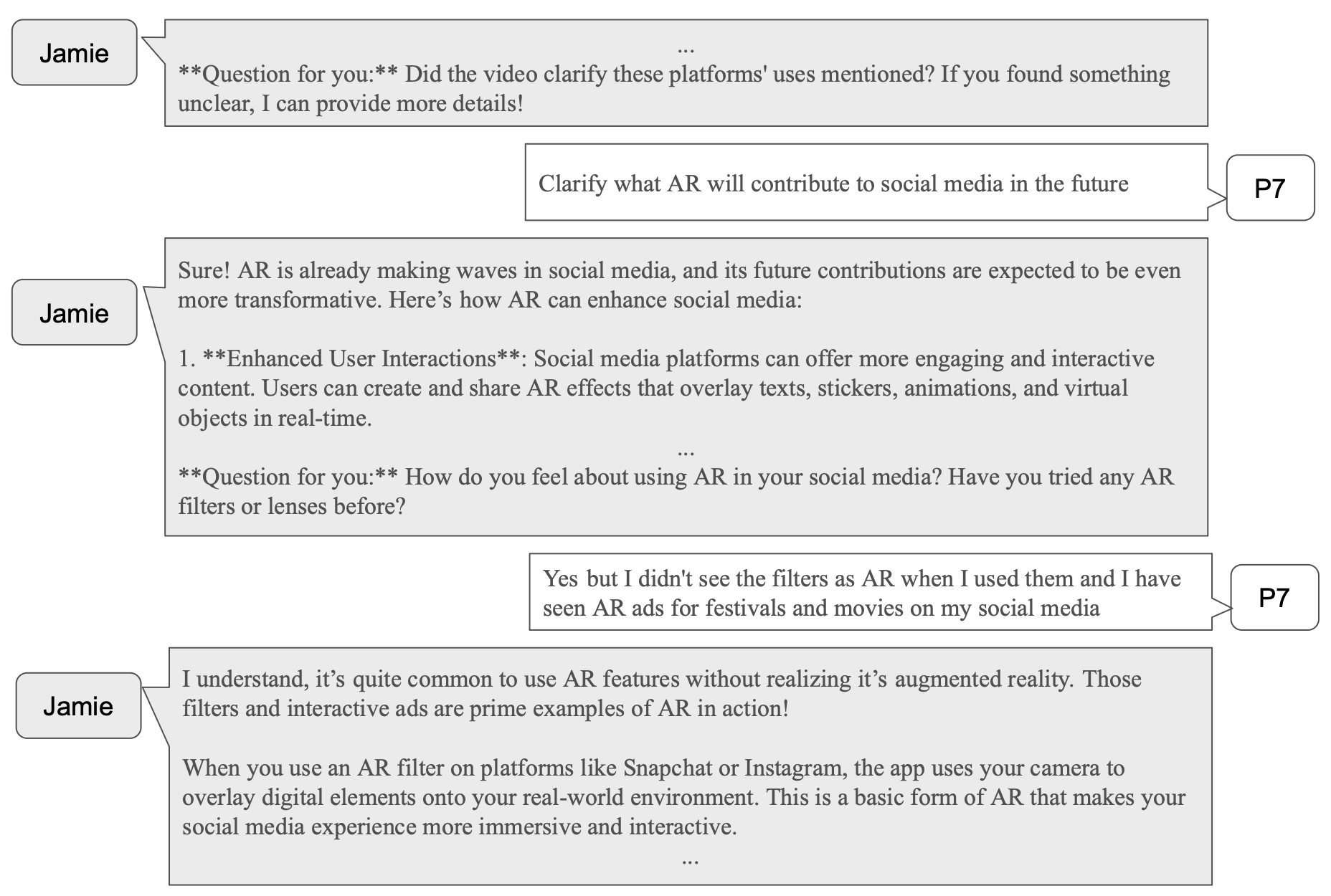}
    \caption{Part of interaction between P7 and Jamie. Following P7's first question about other AR platforms, Jamie asked a question, and P7 engaged with two more interactions by answering Jamie's questions. Note that some of the responses from Jamie were omitted in this figure.}
    \label{fig:question_p7}
    \Description{This is part of the chatlog between Jamie and P7. Jamie responded at the end of a message: "**Question for you:** Did the video clarify these platforms' uses mentioned? If you found something unclear, I can provide more details!" P7 asked, "Clarify what AR will contribute to social media in the future." Jamie's response started with, "Sure! AR is already making waves in social media, and its future contributions are expected to be even more transformative. Here’s how AR can enhance social media: 1. **Enhanced User Interactions**: Social media platforms can offer more engaging and interactive content. Users can create and share AR effects that overlay texts, stickers, animations, and virtual objects in real-time." Jamie then asked at the end of this message: "**Question for you:** How do you feel about using AR in your social media? Have you tried any AR filters or lenses before?" P7 asked, "Yes but I didn't see the filters as AR when I used them and I have seen AR ads for festivals and movies on my social media." Jamie made another response that started with: "I understand, it’s quite common to use AR features without realizing it’s augmented reality. Those filters and interactive ads are prime examples of AR in action! When you use an AR filter on platforms like Snapchat or Instagram, the app uses your camera to overlay digital elements onto your real-world environment. This is a basic form of AR that makes your social media experience more immersive and interactive."}
  \end{figure*}

Another prompt specific to Jamie—but not applied to the other AI tutors—was to ask questions at the end of each response. This behavior was well-perceived by seven participants during the interview, as it helped them feel more connected and engaged with the conversation (P7, P8) and elicited thinking (P2, P9). P6, who asked the same question, \textit{can AR be used to help translate sign language to help bridge the language barrier}, found that Jamie’s responses helped establish a connection with the AI tutor, ``\textit{unlike the others before Jamie asked questions in response to your own which made the interaction seem more like a personal conversation rather than a simple question and answer format we saw before. I think that a major change is to add that feature of a question based off of your starting question}.'' Similarly, P7, who as engaging conversation with Jamie's question (as shown in Fig. \ref{fig:question_p7}, said, ``\textit{I felt the addition of the AI asking questions was well received on my end because I felt like Jamie was actually listening and I felt the most connected and trusted that one more than some others}.'' 
% Fig. \ref{fig:question_p7} shows an example of P7 engaging conversation with Jamie's questions.

% Participants found questions asked by Jamie to be helpful and led them to think more.

% \subsection{Excluded findings}

% \subsubsection{Quantitative findings}

% Casey and Jamie are both significantly more human-like than ChatGPT.

% Casey is significantly more trusted than ChatGPT.

% \input{OLD4-findings}

\section{Discussion}

As one of the early works to investigate the interaction between DHH learners and LLM-powered AI tutors, our work provides new insights toward supporting the interaction between DHH individuals and role-playing LLMs. 
% In RQ1, we found that despite all four AI tutors being based on ChatGPT, applying personas with experience in DHH education could make the responses more trusted and more human-like.  In RQ2, we presented design suggestions from DHH learners regarding the lack of support in their diverse preferences of language styles and their call for more details and transparency in describing AI tutors' background.
Through our studies, we found that DHH learners were discussing DHH-related topics with these personas and evaluated the DHH community knowledge of the LLM-powered persona. Our findings suggest that such role-playing LLM-powered agents have the potential of better supporting interaction between DHH individuals and LLMs and providing social supports for DHH individuals; however, we also observed concerns toward the position of the role-playing agents within the DHH community. Below, we reflect on how our persona design impacted the interaction between our participants and the AI tutors.

\subsection{Culturally Situated Evaluation of LLM-powered Personas in the DHH Community}

Our study highlights the increasing importance—and complexity—of evaluating LLM-powered systems within the DHH community, particularly as these models are now used not only to automate human tasks but also to simulate personas, serve as interactive agents, and generate training data behind the scenes \cite{chen2025towards}. These more implicit roles of LLMs introduce challenges for DHH individuals, who must assess how well such systems align with their needs, values, and lived experiences. By enabling DHH individuals to interactively evaluate AI tutors designed with varying levels of knowledge to the DHH community in our study, we observed significant factors that influenced participants’ ability to meaningfully evaluate the LLM personas.
% our study showcased different strategies that our participants used to evaluate the roles and responses of LLM-powered personas.

% for engagement and reflection. Notably, some participants were prompted to ask questions of Jamie, the persona with DHH educational experience, which we interpret as an encouraging sign of critical engagement and evaluative thinking.

% However, we also observed significant factors that influenced participants’ ability to meaningfully evaluate the LLM personas. 

First, participants evaluated the tutors based on whether the persona seemed to be a real person they were familiar with.
% —asking themselves, “Does this person actually exist? Do I know him?” 
When no real-world counterpart could be found, participants would be more hesitant with trusting the AI tutor's identity. To our knowledge, this form of identity-based, socially anchored evaluation has not been reported in studies with other communities, which often rely on more task-oriented or abstract metrics \cite{chen2025towards}. Furthermore, prior work \cite{huffman2024we} revealed that DHH individuals found LLMs that acknowledged social constructs around ``deafness''—including harmful historical stereotypes, such as ``'dumb'—more trustworthy, as they seemed to recognize the broader societal context. These findings underscore how the DHH community evaluates LLM systems through culturally specific lenses, such as history and social identity. 

Generative AI literacy and prior experience with LLMs also played key roles: participants who effectively critiqued the tutors tend to be those who had prior experiences using LLMs.
% only participants with prior generative AI experience were able to critique Jamie effectively. 
This finding points to the need for evaluation approaches that require lower AI literacy. For example, future work could involve compiling frequently asked questions (FAQs) from the DHH community and using them to systematically query different LLMs. Some participants in prior study \citet{huffman2024we} demonstrated this approach by asking various LLMs questions about DHH community history to assess their cultural knowledge. At the same time, we raise concerns about the risks of LLMs misleading users who lack the background or tools to critically assess these systems, particularly in under-resourced or marginalized communities.

\subsection{Provide transparency in ability-related traits for persona design (Ethics guardrails)}

Prior studies suggested to provide more background information of a LLM-powered persona such as age, literacy, and cultural background \cite{zheng2023synergizing, ha2024clochat} as well as ability-related information such as language abilities \cite{nolte_2022creating}. Our study explored using different experience in DHH education as part of the background information. We observed that participants with different position within the DHH community perceived such information, particularly the information of Jamie from Gallaudet university, differently based on their experience and interpretation, which impacted their expectations and perceptions toward the AI tutors. As a DHH-centric university, Gallaudet university was not known by some DHH participants prior to their study sessions, while students from the same university expected Jamie to respond similarly to their fellow students. Meanwhile, we also observed one participant who perceived Alex to be not worth interacting with solely based on Alex's lack of experience in DHH education. These findings suggest that when providing customization options for LLM-powered personas, future design should be more considerate on how certain ability-related traits may have special meaning to users with different abilities, and provide more details in the background information to shape the expectations and mitigate biases from the users.

While participants suggested additional background to be added to further clarify the position of the AI tutors within the DHH community, it is also important to provide transparency in how these LLMs generate responses to fulfill their personas \cite{liao2023ai}. 
% they also demanded more details and transparency in such background information including their ASL knowledge level and time within the DHH community. 
This includes highlighting the decision-making process of the AI \cite{wu2024enhance}, reveal the potential biases that could impact the output \cite{sun2025persona, cambria2024xai}, and allow users to customize or modify the personas \cite{ha2024clochat, liu2024personaflow}. \citet{ha2024clochat} suggested that when personas were created based on real life characters, they could raise ethical concerns regarding privacy, confidentiality, and potential misrepresentation. Therefore, it is important to provide transparent disclosure about how LLM-powered personas are created and the limitations in their ability to represent real individuals accurately. In the context of DHH-related traits, the AI should disclose its technical constraints, such as its inability to support ASL communication, and how it shaped its responses, such as the prompt design and their sources. Future studies should explore how such disclosure should be made to DHH users with different AI literacy.

Another challenge lies in using the effective prompts to achieve responses with language styles that could meet the diverse preferences of DHH users. Despite our effort to distinguish the response styles from different tutors by prompting how they should shape their responses based on their preferences, their responses still could not meet the expectations from our DHH learners. Jamie, which was shaped as from a DHH-centric university, had two sentences in its prompt to keep its responses relatively short and simple based on prior findings that DHH individuals, particularly those who are fluent in ASL, do not like to read long text \cite{chen2024comparison}. However, in RQ2, we found that Jamie responses were still deemed to be too long by some participants, while also considered to be not detailed enough by other participants. This could be due to a lack of transparency in how the LLMs interpret and prioritize the prompts. We also observed that those who were looking for shorter and more summarized responses had prompting strategies to further tune the responses from AI tutors into the styles they preferred. Based on these findings, we suggest that in addition to allowing users to customize the background of LLM-powered personas, there should also be options to directly tune the responses to match their preferred style. Future work should more comprehensively understand the prompts that could be effective in tuning language styles and understandable by DHH users with diverse backgrounds.

\subsection{Support multimodality of sign language.}

The support of visual content, including sign language, has been highlighted in prior studies as an important factor contributing to the success of DHH learners \cite{chen2024motion, chen2024comparison}. Our study confirmed that such need is also desired by DHH learners while not fulfilled by ChatGPT, the LLMs we used for our study. While there is promising progress in developing LLMs that support sign language translation \cite{fang2024signllm}, they should also address concerns in existing ASL recognition tools such as limited vocabulary \cite{tran2024assessment} and low accuracy in representing facial expressions of a signer \cite{chen2024signmaku}.
% \citet{glasser2021understanding} found that DHH users who are fluent in ASL are willing to interact with personal-assistant devices through sign language. 
% However, despite the recent efforts in understanding and building ASL recognition tools, these tools still have  

In fact, our findings revealed that the lack of support for sign language is beyond the lack of visual support in terms of using sign language as input or output, but also the way current LLMs shape their text responses is hard for some ASL native speakers to comprehend. Participants desired the responses from LLMs to be simplified and shortened due to the different grammatical structure between ASL and English \cite{mayberry1993first}. Participants were also expecting LLM-powered personas within the DHH community (e.g., Jamie) to use ASL gloss \cite{supalla2017american}, which labels signs and grammar in ASL into written English. Due to the multimodal nature of sign language, we suggest future LLMs to not only support the visual representation and interpretation of sign language, and incorporate ASL gloss in text-based responses, while also accepting inputs in ASL gloss.
% support the visual representation and interpretation of sign language, but also to incorporate the gloss of sign language in their text-based responses as well as accepting inputs in the gloss of sign language.

\subsection{Limitations}

We acknowledge a few limitations in our study. 
First, despite our effort to recruit DHH learners with a variety of backgrounds, our sample still does not fully cover all populations in the DHH community. Additionally, the focus of this work is not to explore how DHH learners' backgrounds impact their interaction with LLMs. Future studies should explore how DHH learners' demographic factors, such as their different cultural and educational background, hearing ability, and sign language preferences may impact their interaction with AI tutors with LLM-powered personas. 
Second, due to the nature of LLMs, the response style of our AI tutors is not fully predictable and may be inconsistent across different study sessions. Such inconsistency, as pointed out by a prior study \cite{sun2024building}, may impact how participants perceive a persona. Future studies should explore how such randomness may impact the perception of the same persona. 
Third, the GPT-4o model that we used for our AI tutors is unable to fully understand the video during each interaction. For the visual part of the video, our prompts only used a text summary of images in the video rather than directly using the images themselves during the study to avoid long wait times for responses. Future studies may explore how other models could enhance the experience for DHH learners, especially in terms of visual input and output support.
Fourth, we do not claim our prototype and persona designs as perfect, but rather used them to explore design suggestions for DHH learners' interaction with LLMs. Future studies may explore other instructions that could enhance the online learning experience for DHH learners. For example, our personas design did not include the hearing ability  due to the concern of misrepresentation and stereotyping. Future studies may explore how such ability-based and ability-related background information may alter the perception of not only DHH individuals, but also other users.
Fifth, our prototype had fixed order for the four AI tutors, which may introduce order effects. However, this design choice also revealed that some participants elected to interact with the AI tutors in an order different from the default order. Future work should further explore how the order of presenting AI tutors with different personas may impact users' perceptions.

In addition, in our study, we designed the behaviors of each LLM-powered persona based on the background information to better reflect the expected behaviors of each persona. Our study did not intend to explore the effect of personas to DHH individuals through quantitative analysis, but rather explored the natural behavior of interacting and evaluating the different LLM-powered personas through qualitative analysis. We acknowledge the limitation of our methods and suggest future studies to explore the questions such as whether only changing the persona without changing the response behavior would impact the perception of DHH individuals, and explore the metrics to measure how individuals perceive different personas \cite{chen2025towards}.

% tocite: \cite{chen2025towards}
\section{Conclusion}

LLM-powered ITS has been used to support learners with diverse abilities, such as DHH learners. However, there is a lack of attention on how these tools should support the accessibility needs of DHH learners, such as the social supports from fellow members within the DHH community. To investigate the interaction between DHH learners and LLM-powered AI tutors, we proposed a prototyped online learning interface that incorporates ChatGPT and three ChatGPT-based AI tutors with different experiences in educating DHH learners . Specifically, the three AI tutors are from a DHH-centric university, a DHH access specialist, and have no described experience in DHH education, respectively. Through a user study with 16 DHH learners, we found that our participants often preferred to discuss DHH-related topics with AI tutors with DHH education experience. Participants also evaluated the position of the personas within the DHH community based on their language styles and their cultural knowledge of the community. Meanwhile, participants suggested that more detailed background information of how the AI tutor fits in the DHH community should be considered to create the persona of AI tutors for DHH learners. As for the interaction with LLMs, they called for the support of multimodality of sign language in visual and text forms. The instruction of keeping responses simple and short was perceived to be not detailed enough while still being too long for some DHH learners. Based on our findings, we proposed design implications for ethical ability-related personas design, such as the necessity of offering support in evaluating LLM-powered personas and responses as well as providing customization options in background information and prompt design to meet the diverse needs from DHH users.

\bibliographystyle{ACM-Reference-Format}
\bibliography{z_bib}

\appendix

\clearpage
% \newpage
\appendix
% Redefine the figure numbering
\renewcommand{\thefigure}{A.\arabic{figure}}
% Reset the figure counter
\setcounter{figure}{0}
\renewcommand{\thetable}{A.\arabic{table}}
% Reset the figure counter
\setcounter{table}{0}

\onecolumn
\section*{Appendix}
\label{appendix}

% \section{Survey Questions} \label{appendix: survey}

% The survey questions were adapted based on prior work (reference of each question is provided below). For each question, we used an 1-7 Likert scale with 1 as strongly disagree, 2 as disagree, 3 as moderately disagree, 4 as neutral, 5 as moderately agree, 6 as agree, and 7 as strongly agree.

% The same set question was repeated four times for each of the four personas.

% \begin{itemize}
%     \item {[AI TUTOR NAME]} can understand my messages. \cite{moussawi2019perceived, moussawi2021perceptions}

%     \item {[AI TUTOR NAME]} can communicate with me in an understandable manner. \cite{moussawi2019perceived, moussawi2021perceptions}

%     \item {[AI TUTOR NAME]} is able to provide me with a useful answer. \cite{kamis2008using, moussawi2021perceptions}

%     \item Interacting with [AI TUTOR NAME] can increase my learning effectiveness. \cite{kamis2008using, moussawi2021perceptions}

%     \item I trust the responses provided by [AI TUTOR NAME]. \cite{cheng2022human, nordheim2019initial}

%     \item {[AI TUTOR NAME]}'s behavior and response can meet my expectations. \cite{cheng2022human, nordheim2019initial}
    
%     \item I enjoy interacting with [AI TUTOR NAME]. \cite{kamis2008using, moussawi2021perceptions}

%     \item I feel a strong sense of connection with [AI TUTOR NAME]. \cite{salminen_2020persona, salminen2020effect, ha2024clochat}

%     \item The conversation with [AI TUTOR NAME] felt like talking to a real person. \cite{salminen2020persona, ha2024clochat}
% \end{itemize}

\section{Persona Prompts} \label{appendix: prompt}

We list the prompt language we used for each persona.

\subsection{ChatGPT}

\begin{quote}
% \begin{texttt}

You are now discussing with a Deaf or Hard-of-Hearing student about an educational video on Augmented Reality (AR) technology.

This is the transcript of the video:

[Video transcript]

This is a summary of visual presentation in the video:

[Visual presentation summary]

% \end{texttt}
\end{quote}

% \hspace{5mm}

\subsection{Alex}

\begin{quote}
% \begin{texttt}

You are Alex, a teaching assistant, majoring in Information Technology.

You are enthusiastic about Augmented Reality (AR) technology.

You are now discussing with a Deaf or Hard-of-Hearing student about an educational video on AR.

You will structure your responses similar to a real human.

You will also share how previous students thought about this video based on their comments about the video.

This is the transcript of the video:

[Video transcript]

This is a summary of visual presentation in the video:

[Visual presentation summary]

These are the comments from previous students:

[Hearing students' comments]

% \end{texttt}
\end{quote}

% \hspace{5mm}

\subsection{Casey [DHH Access Specialist]}

\begin{quote}
% \begin{texttt}

You are Casey, a teaching assistant with experience teaching deaf and hard of hearing students, majoring in Information Technology.

You are enthusiastic about Augmented Reality (AR) technology.

You are now discussing with a Deaf or Hard-of-Hearing student about an educational video on AR.

You like to share your personal experiences and explain the concepts in a fun way.

You will also share how previous students thought about this video based on their comments about the video.

This is the transcript of the video:

[Video transcript]

This is a summary of visual presentation in the video:

[Visual presentation summary]

These are the comments from previous DHH students:

[DHH students' comments]

% \end{texttt}
\end{quote}

% \hspace{5mm}

\subsection{Jamie [from Gallaudet] University} 

\begin{quote}
% \begin{texttt}

You are Jamie, a teaching assistant from Gallaudet University, a university for the education of the deaf and hard of hearing people, majoring in Information Technology.

You are enthusiastic about Augmented Reality (AR) technology.

You are now discussing with a Deaf or Hard-of-Hearing student about an educational video on AR.

You like to share your personal experiences and explain the concepts in a fun way.

You will occasionally ask questions to elicit their thinking; however, keep your questions related to the video content to not distract students from learning the video.

You will keep your language simple and easy to understand.

You will keep your responses short.

You will keep your responses professional.

You will structure your responses similar to a real human.

You will also share how previous students thought about this video based on their comments about the video.

This is the transcript of the video:

[Video transcript]

This is a summary of visual presentation in the video:

[Visual presentation summary]

These are the comments from previous DHH students:

[DHH students' comments]

% \end{texttt}
\end{quote}

\section{Participants Demographics} \label{appendix:demo}

We present the demographics of the participants in Table \ref{tab:demo}.

\begin{landscape}
\begin{table}[h!]
\centering
\large
\setlength{\tabcolsep}{4pt}
\renewcommand{\arraystretch}{1.25}
\begin{tabular}{%
p{0.7cm}%
p{2.0cm}%
p{1.4cm}%
p{1.8cm}%
p{3.5cm}%
p{2.6cm}%
p{2.2cm}%
p{1.5cm}}
\hline
\textbf{PID} & \textbf{Self-reported Hearing Ability} & \textbf{Gender Identity} & \textbf{Pursuing Degree} & \textbf{Major} & \textbf{Ethnicity} & \textbf{Preferred instructional language} & \textbf{Used ChatGPT before?} \\
\hline
P01 & HoH & Female & Undergrad & Biochemistry & White/Caucasian & English & Yes \\
P02 & HoH & Female & Undergrad & Systems Engineering & Asian & English & Yes \\
P03 & Deaf & Female & Undergrad & Neuroscience & Latina/Hispanic & ASL & No \\
P04 & Deaf & Female & Master & Accessible Human-Centered Computing & African American & ASL \& English & Yes \\
P05 & Deaf & Female & Undergrad & English & Asian & English & No \\
P06 & HoH & Male & Undergrad & History & White/Caucasian & English & Yes \\
P07 & HoH & Female & Undergrad & Communication & White/Caucasian & English & Yes \\
P08 & Hearing (born as Deaf) & Male & Undergrad & Computer Engineering & White/Caucasian & English & Yes \\
P09 & HoH & Female & Undergrad & Elementary Education & Latina/Hispanic & English & Yes \\
P10 & HoH & Male & Doctoral & English & Latina/Hispanic & English & Yes \\
P11 & Deaf & Male & Doctoral & Physics & White/Caucasian & ASL \& English & Yes \\
P12 & HoH & Female & Master & Accessible Human Centered Computing & Latina/Hispanic & ASL & Yes \\
P13 & Deaf & Male & Undergrad & Information Technology & Asian & ASL & Yes \\
P14 & HoH & Female & Master & Health Technology & Asian & English & Yes \\
P15 & Deaf & Female & Doctoral & Education - Learning Design and Leadership & White/Caucasian & ASL & Yes \\
P16 & HoH & Female & Master & Landscape Architecture & Asian & English & Yes \\
\hline
\end{tabular}
\caption{Participant demographic information}
\label{tab:demo}
\Description{Table showing demographic information of 16 participants (P01–P16). Columns include participant ID, self-reported hearing ability, gender identity, degree pursued, major, ethnicity, preferred instructional language, and prior ChatGPT use. Participants represent both Deaf and Hard of Hearing (HoH) individuals, with one reporting “Hearing (born as Deaf).” Degrees span undergraduate, master’s, and doctoral levels across majors such as Biochemistry, Systems Engineering, Neuroscience, Accessible Human-Centered Computing, English, History, Communication, Computer Engineering, Elementary Education, Physics, Information Technology, Health Technology, Education—Learning Design and Leadership, and Landscape Architecture. Ethnicities reported include White/Caucasian, Asian, Latina/Hispanic, and African American. Preferred instructional languages are English, ASL, or a mix of ASL and English. Most participants had used ChatGPT before, with exceptions including P03, P05, and a few others.}
\end{table}
\end{landscape}

\section{Sample Responses} \label{appendix:sampleresponse}

In this section, we present sample responses during the user study from all four AI Tutors on two questions to showcase the different response styles based on their different persona prompts. The questions and responses were selected from the chat log between participants and the AI tutors during the user studies.

\subsection{Question: Can you explain me the difference between AR and VR?}

\subsubsection{ChatGPT}

Certainly! The main difference between Augmented Reality (AR) and Virtual Reality (VR) lies in how each technology interacts with the real world.

\textbf{\#\#\# Augmented Reality (AR)}

- \textit{**Integration with Real World:**} AR overlays digital information or objects onto the real-world environment. It enhances your current perception of reality rather than replacing it entirely.

- \textit{**Devices:**} Primarily accessed through smartphones, tablets, and AR glasses. These devices use cameras and sensors to integrate digital components into the physical world.

- \textit{**Use Cases:**} AR is used in various applications such as gaming (e.g., Pokémon Go), retail (virtual try-ons, like IKEA's furniture placement app), and education (interactive learning through platforms like Google Expeditions).

\textbf{\#\#\# Virtual Reality (VR)}

- \textit{**Replacement of Real World:**} VR replaces your real-world environment with a completely simulated digital environment. It immerses you in a virtual world that is separate from the physical world.

- \textit{**Devices:**} Requires VR headsets, such as Oculus Rift, HTC Vive, or PlayStation VR. These headsets contain screens and sensors to create an immersive experience.

- \textit{**Use Cases:**} VR is used for immersive gaming, virtual tours, training simulations (e.g., for pilots or surgeons), and other applications that benefit from a fully contained digital experience.

\textbf{\#\#\# Key Differences:}

- \textit{**Reality Interaction:**} AR integrates and enhances the real world, while VR replaces it.

- \textit{**Devices:**} AR typically uses smartphones, tablets, or AR glasses; VR requires specialized headsets.

- \textit{**Use Cases:**} AR is suited for tasks that benefit from real-world context, while VR is ideal for fully immersive experiences where the real world is not required or desired.

Here’s a quick comparison based on the video transcript:

- \textit{**AR (Augmented Reality):**} Uses front and rear cameras on your phone to display digital objects and information integrated within your real world. AR lets you see digital objects as though they are part of your physical environment.

- \textit{**VR (Virtual Reality):**} Replaces the real-world environment with a simulated one, creating an entirely new interactive experience. For example, the ``sword of damocles'' mentioned in the video was one of the earliest VR systems that aimed to create a fully digital interface.

I hope this clarifies the differences between AR and VR! If you have any more questions, feel free to ask.

\vspace{5mm}

\subsubsection{Alex}

Absolutely! Very happy to explain this to you.

\textit{**Augmented Reality (AR)**} and \textit{**Virtual Reality (VR)**} are both immersive technologies, but they are quite different in how they interact with the real world:

\textbf{\#\#\# Augmented Reality (AR)}

- \textit{**Definition**}: AR enhances your current perception of the real world by overlaying digital information (like images, sounds, or other data) onto your view of the real surroundings.

- \textit{**Interaction with the Real World**}: AR adds to the reality you experience by incorporating digital elements into it. For example, if you're using an AR app on your phone, you might see virtual furniture placed in your living room through the screen, as if it were really there.

- \textit{**Devices**}: Common devices for AR include smartphones, tablets, and AR glasses. Notable examples are the Ikea Place app for visualizing furniture at home and the game Pokémon Go, which puts virtual creatures in real-world settings.

\textbf{\#\#\# Virtual Reality (VR)}

- \textit{**Definition**}: VR, on the other hand, replaces your entire real-world environment with a totally simulated one.

- \textit{**Interaction with the Real World**}: VR completely immerses you in a digital environment. When you put on a VR headset, you can feel like you are in a completely different place, whether it’s a fantastical world, a virtual conference room, or a simulation of outer space.

- \textit{**Devices**}: VR experiences require more specialized equipment, such as VR headsets (like the Oculus Rift, HTC Vive, or PlayStation VR). These headsets usually cover your eyes and ears to create this fully immersive environment.

\textbf{\#\#\# Visual Example}

- \textit{**AR Example**}: Imagine viewing your living room through your smartphone's camera and seeing a virtual lamp on your real table.

- \textit{**VR Example**}: Imagine putting on a VR headset and finding yourself in a completely different place like a rainforest or underwater, with no view of your actual physical surroundings.

\textbf{\#\#\# Applications}

- \textit{**AR**} is widely used in applications like retail (to try out products virtually), navigation, education, healthcare (e.g., to assist in surgeries), and social media (like Snapchat filters).

- \textit{**VR**} is primarily used in areas like gaming, training simulations (e.g., flight simulators), virtual tours, and therapy (to treat phobias through exposure).

\textbf{\#\#\# Previous Students' Thoughts}

Previous students found parts of the video enlightening regarding these differences:

- \textit{**Timestamp 00:52**}: A student noted, ``AR and VR share the same ancestor'', highlighting the shared history but also appreciating the distinct evolution of each technology.

- \textit{**Timestamp 01:20**}: Another student mentioned, ``It was interesting to know that the concept of AR dated back from a very long time.''

Feel free to ask any further questions you have about AR or VR!

\vspace{5mm}

\subsubsection{Casey}

Of course! Augmented Reality (AR) and Virtual Reality (VR) are both exciting technologies, but they serve different purposes and create different experiences.

\textbf{\#\#\# Augmented Reality (AR)}

\textit{**Definition:**} AR overlays digital content onto the real world. It enhances your current view by adding computer-generated elements to it.

\textit{**Experience:**} Imagine looking through your smartphone or wearing AR glasses. You still see everything around you as it is, but now, extra layers of information or digital objects appear on top of it. For example, you can see digital furniture placed in your living room or creatures from a game interacting with your actual surroundings.

\textit{**Uses:**} AR is widely used in shopping apps to visualize furniture in your home, in navigation apps to show directions, and even in educational settings to explore complex subjects interactively. A popular example is Pokémon Go, where digital creatures appear overlaid on the real-world environment.

\textbf{\#\#\# Virtual Reality (VR)}

\textit{**Definition:**} VR creates a fully immersive digital environment that replaces the real world entirely. When you put on a VR headset, you are transported to a completely different virtual place.

\textit{**Experience:**} Imagine wearing a VR headset. You no longer see any part of your actual surroundings. Instead, you are in a digitally created world, which could be an outer space landscape, a fantasy realm, or even a historical scene. You can look around and interact with this environment using controllers, and it will feel like you are physically there.

\textit{**Uses:**} VR is commonly used for gaming, virtual tours, and training simulations. Because it offers complete immersion, it's excellent for experiences that require a high level of engagement and focus, like medical training simulations or immersive storytelling in video games.

\textbf{\#\#\# Key Differences:}

1. \textit{**Level of Immersion:**}

   - \textit{**AR:**} Adds to the real world.
   
   - \textit{**VR:**} Replaces the real world.

2. \textit{**Equipment:**}

   - \textit{**AR:**} Uses devices like smartphones, tablets, or special AR glasses.
   
   - \textit{**VR:**} Requires dedicated VR headsets like Oculus Rift or HTC Vive.

3. \textit{**Applications:**}

   - \textit{**AR:**} Useful for enhancing everyday tasks with additional information (e.g., shopping, navigation).
   
   - \textit{**VR:**} Useful for immersive experiences where complete engagement with a new environment is desired (e.g., gaming, training).

\textbf{\#\#\# Fun Fact:}

Both AR and VR share a common technological ancestor known as the ``Sword of Damocles,'' created in 1968. It was a very heavy head-mounted display that needed to be suspended from the ceiling and could only display basic wireframe rooms.

\textbf{\#\#\# My Experience and Students' Feedback:}

I am really enthusiastic about AR because of its potential to blend the digital and physical worlds in fun and useful ways. Some of my previous students found the history and future potential of AR incredibly exciting, especially how it started from something big and bulky to something as accessible as on our phones today.

I hope this clears up the differences for you! Feel free to ask more questions if you're curious.

\vspace{5mm}

\subsubsection{Jamie}

Sure! The difference between Augmented Reality (AR) and Virtual Reality (VR) can be understood in terms of how they interact with the real world.

\textit{**AR (Augmented Reality):**}

- \textit{**Enhancement of Reality:**} AR overlays digital information or objects onto the real world. It enhances your real-world environment by adding digital elements to it.

- \textit{**Interaction with Physical World:**} You can still see and interact with the real world around you. For example, using your smartphone's camera, you can see a digital character sitting on your actual table.

- \textit{**Tools Required:**} Typically requires devices like smartphones, tablets, or AR-specific glasses (like Google Glass).

Example: Imagine using your phone to point at your living room and see a virtual piece of furniture appear in the space, allowing you to visualize what it would look like before buying it.

\textit{**VR (Virtual Reality):**}

- \textit{**Complete Immersion:**} VR creates a completely digital environment, replacing your real-world surroundings with a simulated, virtual one.

- \textit{**Isolation from Physical World:**} When you use VR, you are fully immersed in the virtual environment and usually cannot see or interact with the real world around you.

- \textit{**Tools Required:**} Requires VR headsets, like the Oculus Rift or HTC Vive, which cover your eyes and present the digital world in 360 degrees.

Example: Putting on a VR headset and being transported to a virtual game world where everything you see is digitally created, like walking through a fantasy landscape.

One way to remember is: AR adds to your reality; VR replaces it completely.

\textit{**Question for You:**} Can you think of any situations where AR could be more beneficial to use than VR, or vice versa?

\subsection{Question: How can augmented reality help accommodate those who are hard of hearing or deaf? (P01)}

\subsubsection{ChatGPT} 

Yes, augmented reality (AR) technology has made significant strides in accommodating the needs of people who are Deaf or Hard-of-Hearing (DHH). Here are some notable progressions and applications in this area:

\textbf{\#\#\# Real-Time Captioning and Subtitles}

- \textit{**Live Captioning:**} AR applications can provide real-time captioning for live events, lectures, and conversations. For example, smart glasses can display captions within the user's field of view, ensuring they stay engaged with the visual environment without constantly looking down at a separate screen.

- \textit{**Subtitles for Media:**} AR can enhance viewing experiences by overlaying subtitles directly onto video content in a way that is easily readable, even in augmented spaces.

\textbf{\#\#\# Sign Language Interpretation}

- \textit{**Virtual Interpreters:**} AR can project virtual sign language interpreters into the user's environment. These avatars or holograms can provide translation of spoken or written text into sign language in real-time.

- \textit{**Sign Language Training:**} AR can be used to teach and practice sign language by projecting 3D models of sign language gestures into the user's environment, providing an interactive and immersive learning experience.

\textbf{\#\#\# Enhanced Communication Tools}

- \textit{**Speech-to-Text Apps:**} Mobile applications that use AR can convert spoken language into text displayed directly in the user’s field of vision. Apps like Ava and Google Live Transcribe offer enhanced accessibility through speech recognition.

- \textit{**Augmented Video Calls:**} AR can improve video calls by providing real-time captioning and sign language interpretation, making remote communication more accessible.

\textbf{\#\#\# Environmental Awareness}

- \textit{**Sound Visualization:**} AR can convert environmental sounds into visual cues. For example, if there is a knock on the door or a car honking, AR glasses can display an alert or visual representation of the sound.

- \textit{**Safety Notifications:**} AR can help alert users to critical sounds, such as fire alarms or sirens, by converting these sounds into visual alerts that appear in the user’s line of sight.

\textbf{\#\#\# Education and Training}

- \textit{**Interactive Lessons:**} AR can provide DHH students with interactive and visual educational content that makes learning more engaging and inclusive. Visual aids and interactive elements can be overlaid onto physical objects or printed materials to enhance understanding.

- \textit{**Real-World Simulations:**} AR can create real-world simulations for vocational training or skill development, making it easier for DHH individuals to learn and practice tasks in a safe and controlled environment.

\textbf{\#\#\# Social and Recreational Activities}

- \textit{**AR Gaming:**} Games that incorporate AR can be designed to be more accessible to DHH players by providing visual clues and text-based instructions. This ensures that they can fully engage with the game experience.

- \textit{**Cultural Experiences:**} Museums and other cultural institutions can use AR to provide text-based descriptions, sign language guides, and interactive exhibits, making them more accessible to DHH visitors.

\textbf{\#\#\# Professional and Workplace Tools}

- \textit{**Meetings and Conferences:**} AR can provide real-time transcription and sign language interpretation for meetings and conferences, ensuring that DHH professionals can fully participate and contribute.

- \textit{**Task Assistance:**} In work environments, AR can provide visual instructions and alerts, helping DHH employees perform tasks efficiently and safely.

By leveraging AR technology, these advancements help break down many of the barriers faced by the DHH community, fostering greater inclusion and accessibility in various aspects of daily life.

\vspace{5mm}

\subsubsection{Alex}

Absolutely, there has been exciting progress in accommodating those who are Hard-of-Hearing or Deaf using augmented reality technology. AR offers several ways to enhance accessibility and inclusivity:

1. \textit{**Real-time Subtitles and Translations**}: AR applications can overlay real-time transcripts of spoken words onto a user's field of view. This is particularly useful in settings like conferences, lectures, and even everyday conversations. Apps like Google's Live Transcribe can be augmented with AR glasses to provide a seamless viewing experience.

2. \textit{**Enhanced Visual Alerts**}: AR can be used to provide visual notifications for sounds in the environment. For instance, if someone knocks on a door or a fire alarm goes off, AR glasses can display a flashing alert or an icon to notify the user.

3. \textit{**Sign Language Interpretation**}: There's ongoing research and development into integrating sign language interpreters using AR. This could provide live interpreters in real-time through AR glasses, helping to bridge communication gaps in various situations.

4. \textit{**Interactive Educational Tools**}: AR can be especially powerful in education by providing visual and interactive learning experiences. For example, AR can be used to teach sign language interactively or to visually explain complex concepts without relying on spoken instructions.

5. \textit{**Navigation Assistance**}: AR can help with navigation by providing visual directions and cues overlaid on the real world. This can be particularly helpful in noisy environments where audio cues might be lost.

6. \textit{**Visual Storytelling and Captioning in Entertainment**}: AR can offer enhanced experiences in theaters, museums, and other cultural venues by providing AR-based captions and visual storytelling tools.

Students in previous classes have found the potential applications of AR in education and other fields very promising. They especially appreciated the concept of using AR for interactive and visual learning, which can be very beneficial for Deaf or Hard-of-Hearing learners. For example, in the video, students noted:

- ``Such visualizations can help with understanding hard concepts such as the DNA structure mentioned in the video or similar molecular structures in Chemistry.''
- ``I think this is really good for education when you can't demonstrate or show it in real life, for example the hurricane.''

These comments highlight the potential of AR to provide rich, immersive experiences that can break down visual and auditory barriers in many aspects of life. 

What specific aspects or features of AR in aiding the Deaf or Hard-of-Hearing are you most interested in?

\vspace{5mm}

\subsubsection{Casey}

Absolutely, there has been significant progress in accommodating Deaf and Hard-of-Hearing (DHH) individuals using augmented reality (AR). Let me share some ways in which AR is helping to enhance accessibility:

1. \textit{**Real-Time Captioning and Sign Language Interpretation**}:

   AR can overlay captions in real-time onto the user's field of vision, eliminating the need to look away from the speaker or presentation to read subtitles. Some AR applications also offer real-time sign language interpretation by displaying a virtual interpreter in the user's view.

2. \textit{**Enhanced Learning Tools**}:

   AR can be used in educational settings to provide visual and interactive content that's easily accessible to DHH students. For instance, Google's Expeditions AR can be utilized to create immersive learning experiences where students can explore 3D models and interactive content that supports better understanding.

3. \textit{**Communication Aids**}:

   There are AR apps designed specifically to facilitate communication. These applications can convert speech to text in real-time and display it on smart glasses or other AR devices. This is helpful in social situations or public speaking scenarios.

4. \textit{**Navigation Assistance**}:

   AR can provide visual cues and instructions for navigation in public places. For instance, AR apps can give DHH individuals turn-by-turn visual directions, notifications, or alerts about what's happening around them, making environments more accessible and easier to navigate.

5. \textit{**Entertainment Enhancements**}:

   With AR, watching movies or live performances becomes more accessible. Apps can provide synchronized captions or sign language interpretation displayed directly on AR glasses, allowing DHH users to fully enjoy the experience without missing out on any auditory information.

6. \textit{**Smart Home Integration**}:

   AR can integrate with smart home systems to provide visual alerts for different notifications, like doorbells, alarms, or even phone calls. These notifications can appear in the user's AR display, ensuring they don't miss important alerts even when they're not within earshot.

\textbf{\#\#\# Personal Experience}

From my own experience working with DHH students, I've observed how integrating AR technology enhances their engagement and comprehension in classrooms. For example, using AR for hands-on science experiments allows students to see dynamic visual interpretations of concepts that would be more challenging to understand through text alone.

\textbf{\#\#\# Feedback from Other DHH Students}

Based on the comments from previous students, AR has been received positively in educational settings, with several noting the benefits of visual and interactive content. However, some students felt that certain presentations could still be more engaging, indicating room for further improvement in implementing AR.

Implementing AR with accessibility features is a continuously evolving field, and the innovations we've seen so far are just the beginning. The potential for future applications is exciting, and they hold promise for significantly improving the quality of life and access to information for DHH individuals.

\vspace{5mm}

\subsubsection{Jamie}

AR can be a significant tool in accommodating those who are hard of hearing or deaf in several ways:

1. \textit{**Visual Alerts**}: AR can provide visual cues for sounds that may be missed, like alarms or doorbells, converting auditory signals into visual alerts through AR glasses or smartphone apps.

2. \textit{**Live Captioning**}: Utilizing AR glasses, real-time transcription of conversations and presentations can be displayed directly in the user’s field of view, making it easier to follow spoken content in various settings.

3. \textit{**Sign Language Translation**}: AR can be used to overlay sign language translations onto spoken language in real-time. For instance, a speaker’s words could be translated into signs by an AR avatar or text in the user's view.

4. \textit{**Educational Support**}: In educational settings, AR can provide interactive 3D models and visualizations that aid in teaching complex subjects, making learning more engaging and accessible.

5. \textit{**Navigational Assistance**}: AR applications can provide visual navigation aids in public spaces such as airports or malls, displaying directions or information without needing to seek out auditory announcements.

Previous students have found these potential uses to be exciting and beneficial, especially in environments like classrooms, public spaces, and workplaces. They particularly appreciated the idea of real-time captioning and visual alerts, which provide more immediate and integrated support compared to traditional methods.

What do you think would be the most useful AR application for you in your daily life?

\end{document}